\documentclass[%
 aip,
 amsmath,amssymb,
secnumarabic,%
 reprint,%
]{revtex4-1}

\usepackage[usenames, dvipsnames]{color}
\definecolor{myyellow}{RGB}{255, 255, 192}
\definecolor{mygreen}{RGB}{192, 255, 192}
\definecolor{myblue}{RGB}{192, 192, 255}

\usepackage{gensymb}
\usepackage{colortbl}
\usepackage{amsmath,amssymb,color}
\usepackage{mathrsfs}
\usepackage{upgreek}
\usepackage{dcolumn}% Align table columns on decimal point
\usepackage{graphicx}% Include figure files
\usepackage{docs}%
\usepackage{bm}%
\usepackage[colorlinks=true,linkcolor=blue]{hyperref}%
\usepackage[usenames,dvipsnames]{xcolor}
\usepackage{float}
\usepackage[utf8]{inputenc}
\usepackage[T1]{fontenc}
\usepackage{mathptmx}
\usepackage{etoolbox}
\usepackage{placeins}
\expandafter\ifx\csname package@font\endcsname\relax\else
 \expandafter\expandafter
 \expandafter\usepackage
 \expandafter\expandafter
 \expandafter{\csname package@font\endcsname}%
\fi
\providecommand{\abs}[1]{\lvert#1\rvert}

\makeatletter
\def\@email#1#2{%
 \endgroup
 \patchcmd{\titleblock@produce}
  {\frontmatter@RRAPformat}
  {\frontmatter@RRAPformat{\produce@RRAP{*#1\href{mailto:#2}{#2}}}\frontmatter@RRAPformat}
  {}{}
}%
\makeatother
\begin{document}

\preprint{AIP/123-QED}

\title{Extensional rheometry of mobile fluids. Part I: OUBER, an optimized uniaxial and biaxial extensional rheometer}

%\title[Journal of Rheology]{Extensional rheometry of mobile fluids. Part I: OUBER, an optimized uniaxial and biaxial extensional rheometer}
% Force line breaks with \\
\author{Simon J. Haward}
 \email{simon.haward@oist.jp.}
 \affiliation{Okinawa Institute of Science and Technology, Onna, Okinawa 904-0495, Japan.}%Lines break automatically or can be forced with \\
%\altaffiliation[Also at ]{Okinawa Institute of Science and Technology, Onna, Okinawa 904-0495, Japan.}

\author{Francisco Pimenta}%
\affiliation{Departamento de Engenharia Qu\'{i}mica, ALiCE, CEFT, Faculdade de Engenharia da Universidade do Porto, Rua Dr. Roberto Frias, 4200-465 Porto, Portugal.}%

\author{Stylianos Varchanis}
\affiliation{Okinawa Institute of Science and Technology, Onna, Okinawa 904-0495, Japan.}%

\author{Daniel W. Carlson}
\affiliation{Okinawa Institute of Science and Technology, Onna, Okinawa 904-0495, Japan.}%

\author{Kazumi Toda-Peters}
\affiliation{Okinawa Institute of Science and Technology, Onna, Okinawa 904-0495, Japan.}%

\author{Manuel A. Alves}
\affiliation{Departamento de Engenharia Qu\'{i}mica, ALiCE, CEFT, Faculdade de Engenharia da Universidade do Porto, Rua Dr. Roberto Frias, 4200-465 Porto, Portugal.}%

\author{Amy Q. Shen}
\affiliation{Okinawa Institute of Science and Technology, Onna, Okinawa 904-0495, Japan.}%

\date{7 April 2022}%
\revised{\today}%

\begin{abstract}
We present a numerical optimization of a ``6-arm cross-slot'' device, yielding several three-dimensional shapes of fluidic channels designed to impose close approximations to ideal uniaxial (or biaxial) stagnation point extensional flow under the constraints of having four inlets and two outlets (or two inlets and four outlets) and Newtonian creeping flow conditions. Of the various numerically-generated geometries, one is selected as being most suitable for fabrication at the microscale, and numerical simulations with the Oldroyd-B and Phan-Thien and Tanner models confirm that the optimal flow fields in the chosen geometry are observed for both constant viscosity and shear thinning viscoelastic fluids. Fabrication of the geometry, which we name the optimized uniaxial and biaxial extensional rheometer (OUBER), is achieved with high precision at the microscale by selective laser-induced etching of a fused-silica substrate. Employing a viscous Newtonian fluid with a refractive index matched to that of the optically transparent microfluidic device, we conduct microtomographic-particle image velocimetry in order to resolve the flow field at low Reynolds number ($<0.1$) in a substantial volume around the stagnation point. The flow velocimetry confirms the accurate imposition of the desired and predicted flows, with pure extensional flow at an essentially uniform deformation rate being applied over a wide region around the stagnation point. In Part II of this paper [Haward \textit{et al.}, J. Rheol. submitted (2023)], pressure drop measurements in the OUBER geometry will be used to assess the uniaxial and biaxial extensional rheometry of dilute polymeric solutions, in comparison to measurements made in planar extension using an optimized-shape cross-slot extensional rheometer (OSCER, Haward et al, Phys. Rev. Lett., 2012).   
\end{abstract}

\maketitle

\section{\label{intro}Introduction}

Almost all flows of practical importance are comprised of both shearing and extensional kinematics. Prominent examples include flows through contractions or expansions, around obstacles and through branching junctions. Simple Newtonian fluids can be fully characterized by knowledge of their shear viscosity $\eta$ alone, since their extensional viscosity is known to also be constant and equal to $3 \eta$, $4 \eta$, or $6 \eta$ for uniaxial, planar, or biaxial extension, respectively, where the coefficients 3, 4, and 6 are the respective Trouton ratio, $\text{Tr}$. \cite{Trouton1906,Petrie2006} By contrast, for viscoelastic fluids such as polymeric solutions and melts, which are widely present in industrial and biological processes, the situation is very different. Here, for sufficiently high strain rates, extensional flows are very effective at unraveling and orienting polymer chains. \cite{DeGennes1974,Hinch1974,Keller1985,Larson1989,Perkins1997} Due to the entropic elasticity of the polymer, causing it to resist deformation, the hydrodynamically-forced unraveling results in a non-linear increase in the elastic tensile stress difference, and hence the extensional viscosity, with the nature of the increase being unknown \emph{a priori} due to dependence on the fluid properties (e.g., polymer concentration, molecular weight, and extensibility). \cite{Tirtaatmadja1993,James1994,James1995,Morrison} For viscoelastic fluid flows, even localized regions of extensional kinematics within the flow field can have a dominant impact on the macroscopic flow behavior. \cite{Morrison} For this reason, the quantitative characterization of the extensional viscosity of viscoelastic fluids is essential to enable a fully descriptive prediction of their behavior in arbitrary flow fields. \cite{Barnes1989,Macosko1994} Unfortunately, the measurement of extensional viscosity is nontrivial, with a key challenge being to generate an extensional flow field that is both persistent and spatially uniform. \cite{Petrie2006}

\begin{figure}[t]
\begin{center}
\includegraphics[scale=0.5]{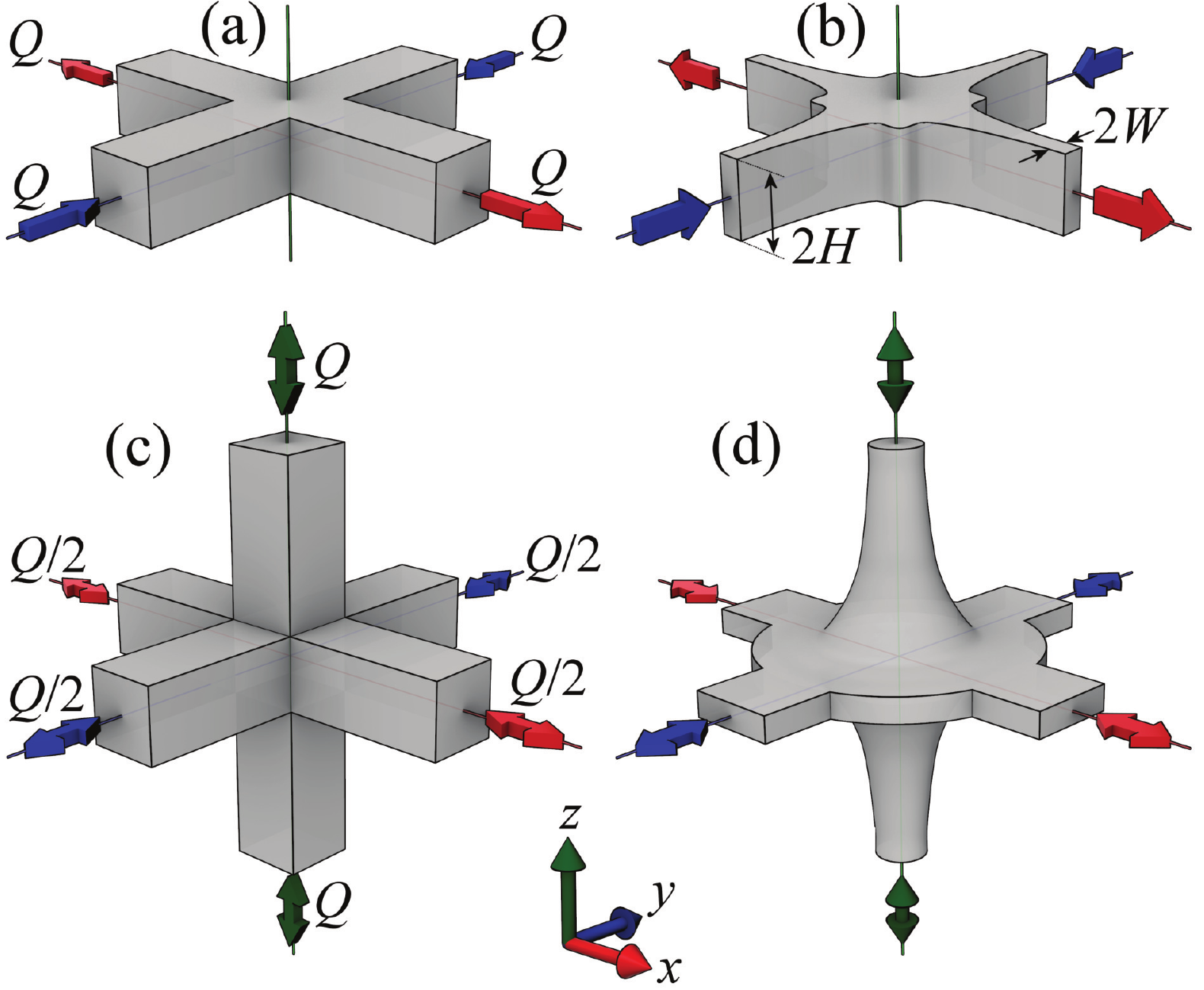}
\caption {Schematic illustrations of (a) standard cross-slot geometry for approximating planar elongation, (b) an optimized shape cross-slot extensional rheometer (OSCER) device, (c) a 6-arm cross-slot for approximating to uniaxial and biaxial elongation, (d) a first guess at the form of an optimized shape 6-arm cross-slot with a pair of opposing circular inlet/outlet channels oriented along $z$ and four planar outlet/inlet channels in the $z=0$ plane. 
}
\label{schematic}

\end{center}
\end{figure}

The potential of stagnation point flows for extensional rheometry has long been recognized, with the cross-slot device (Fig.~\ref{schematic}(a)) being a relevant example. The cross-slot is a simple geometry, easily fabricated at microscale dimensions in order to obviate inertia with even low viscosity fluid samples. It consists of two oppositely-facing rectangular inlet channels (aligned with $y$ in Fig.~\ref{schematic}(a)) joined at right angles to two oppositely-facing rectangular outlet channels (aligned with $x$). By imposing an equal volumetric flow rate $Q$ through each of the four channels, an approximation to planar extensional flow is generated (ideally described by a rate-of-strain tensor $\textnormal{\bf{D}}$ with the only non-zero components being $D_{xx} = -D_{yy} = \dot \varepsilon$, where $\dot \varepsilon$ is the strain rate). Crucially, there exists a free stagnation point at the center of symmetry of the cross-slot geometry where the flow velocity is zero. Hence, $\dot\varepsilon$ ($\propto Q$) is applied persistently allowing strain to accumulate infinitely and any deformation of the microstructure (e.g., polymer), to reach a steady state. The steady state value of the tensile stress difference $\Delta \sigma(\dot\varepsilon)$ can be estimated by appropriate measurements of the pressure drop across an inlet and an outlet of the device,~\cite{Haward2016} with the planar extensional viscosity given by $\eta_P = \Delta\sigma / \dot\varepsilon$.

A problem with the ``standard'' cross-slot device as depicted in Fig.~\ref{schematic}(a), is that the flow field is not homogeneous. The approximation to planar extension is only valid arbitrarily close to the stagnation point, and a given applied $Q$ does not provide a spatially uniform value of $\dot\varepsilon$. By combining a finite-volume flow solver with an automatic mesh generator and an optimizer, Alves (2008) iteratively modified the two dimensional (2D) profile of the cross-slot in the region connecting the inlets and the outlets in order to obtain an optimal approximation to ideal planar elongation. \cite{Alves2008,Haward2012c} The resulting Optimized-Shape Cross-slot Extensional Rheometer (OSCER, Fig.~\ref{schematic}(b)) imposes an almost homogeneous planar elongation over a region spanning $15W$ about the stagnation point, where $W$ is the characteristic channel half-width. \cite{Alves2008,Haward2012c} Since homogeneity is also required through the neutral $z$-direction, the flow should be 2D, so experimentally the OSCER geometry requires a high aspect ratio $H/W \approx 10$, where $H$ is the channel half-height (see Fig.~\ref{schematic}(b)). \cite{Haward2012c} The OSCER geometry has proven useful for characterizing the extensional rheology and flow behavior of a variety of viscoelastic fluids. \cite{Haward2012c,Haward2013b,Haward2016c}

Recently, advancements in three-dimensional (3D) microfabrication methods have motivated the development of a microfluidic 6-arm cross-slot (see Fig.~\ref{schematic}(c)). \cite{Afonso2010,Haward2019b} Such a device can be operated in two modes. By injecting fluid at volumetric rate $Q/2$ along the two pairs of opposed inlets aligned with $x$ and $y$, and withdrawing fluid at volumetric rate $Q$ along the opposed outlets aligned with $z$, an approximation to uniaxial extension ($D_{xx} = D_{yy} = -\dot\varepsilon/2$, $D_{zz} = \dot\varepsilon$) is obtained. By reversing the flow in each channel, an approximation to biaxial extension ($D_{xx} = D_{yy} =  \dot\varepsilon_B$, $D_{zz} = -2 \dot\varepsilon_B$) is obtained. Note that the subscript ``$B$'' on $\dot\varepsilon$ in the case of biaxial extension is to conform to established Society of Rheology notation,~\cite{Meissner1982,Dealy1984,Petrie1984,Dealy1995,Petrie1990} and distinguishes from the case where biaxial extension is considered as uniaxial compression, with $D_{xx} = D_{yy} =  \dot\varepsilon/2$, $D_{zz} = -\dot\varepsilon$.~\cite{Dealy1984,Bird} The 6-arm cross-slot generates a stagnation point at its center and has been described as a microfluidic analog to the opposed jets device. \cite{Frank1971,Fuller1980,Haward2019b} However, microscale fabrication solves the problem of inertia encountered in the classical opposed jets.\cite{Schunk1990,Dontula1997} Also, similarly to the standard cross-slots, there is a possibility to evaluate $\Delta \sigma(\dot\varepsilon)$ (or $\Delta \sigma(\dot\varepsilon_B)$) by appropriate pressure drop measurements. As such, the microfluidic 6-arm cross-slot device possesses some promising attributes for use as a uniaxial and biaxial extensional rheometer for low viscosity ``mobile'' fluids. A question that arises is whether the 3D geometry of the 6-arm cross-slot can be optimized in a way similar to the standard cross-slot device in order to obtain more homogeneous uniaxial and biaxial elongation.

Uniaxial and biaxial extension are kinematically the reverse of each other; uniaxial extension can equally be described as biaxial compression, while biaxial extension can be called uniaxial compression. Such flows have axisymmetry, with extension (compression) along an axis normal to the compressional (extensional) plane. With these considerations in mind, and given the practical constraints of a device with 4 (2) inlets and 2 (4) outlets, a first guess of the form of an optimized 6-arm cross-slot is shown in Fig.~\ref{schematic}(d). In such a device, to generate uniaxial (biaxial) extensional flow there are two opposing circular outlets (inlets) connected to four planar inlets (outlets). The inlets and outlets are connected by a nominally hyperbolic and axisymmetric shape, the precise optimal form of which must be determined.

In this work, we first perform a numerical shape optimization procedure on the 6-arm cross-slot resulting in a number of 3D geometries that provide very close approximations to ideal uniaxial and biaxial extension under Newtonian creeping flow conditions. Selecting one of the optimal geometries for potential fabrication, we confirm its suitability for use with fluids of complex rheology by performing numerical simulations of the flow field with some commonly-used viscoelastic fluid models. Next, we employ the 3D microfabrication technique of selective laser-induced etching (SLE) to fabricate a glass microfluidic geometry according to the selected design. Micro-tomographic particle image velocimetry ($\upmu$-TPIV) with a Newtonian fluid in the device, which we call the Optimized Uni- and Biaxial Extensional Rheometer (OUBER), is used to demonstrate that the intended velocity fields are indeed imposed over a substantial volume of the device, generating a controlled and steady extensional rate. Our ultimate aim with this research program is to undertake a comparison between the uniaxial, planar and biaxial extensional rheology of viscoelastic fluids. It should be noted that although uniaxial and biaxial extension are the kinematic reverse of each other, the two flows are expected to result in significantly different polymer unraveling dynamics. \cite{Petrie2006}  Accordingly, contrasting extensional rheology should result, although there is little by way of convincing evidence from either experiment or theory to confirm or refute this. \cite{Petrie1984,Jones1987,Petrie1990,Kwan2001,Shogin2021}

The remainder of the paper is organized as follows. In Sec.~\ref{NumOpt}, we describe the numerical optimization of the 6-arm cross-slot geometry and we simulate viscoelastic flows in one of the resulting optimal geometries most suited to experimental fabrication. In Sec.~\ref{ExpMeth}, we provide a detailed description of the OUBER device fabrication, and describe our flow measurement methods. The results of our Newtonian flow experiments in the OUBER device are presented and discussed in Sec.~\ref{Res}, and we draw our conclusions in Sec.~\ref{SumCon}.

\section{Numerical Optimization of the 6-arm Cross-Slot}
\label{NumOpt}

In this section we will discuss the numerical optimization of the 6-arm cross-slot device, originally proposed by Afonso \textit{et al}~\cite{Afonso2010} and recently fabricated experimentally by Haward \textit{et al}. \cite{Haward2019b} To date, numerical optimizations of microfluidic device geometries have mostly been performed on 2D planar flow geometries, with the focus on achieving homogeneous extensional rates.\cite{Alves2008,Galindo2014,Zografos2016,Pimenta2018,Zografos2019} Numerical optimization of a 3D flow geometry has only recently been demonstrated by Pimenta \textit{et al}, who generated an optimized shape axisymmetric contraction-expansion geometry designed to produce a near constant extensional rate along its axis. \cite{Pimenta2020}

In the present work, the numerical optimization of the 6-arm cross-slot (Fig.~\ref{schematic}(c)) follows a procedure similar to that outlined by Pimenta \textit{et al}.~\cite{Pimenta2020} Briefly, the geometry is parametrized with a set of design points that can be moved to deform the wall in order to achieve an extensional flow along a predefined region. The motion of these points is controlled by a derivative-free optimizer (Nomad  v.3.9.12),\cite{Digabel2011} whose goal is to minimize an objective function embodying the difference between the flow in each candidate geometry and a theoretical homogeneous extensional flow. The velocity profiles used to compute the objective function are obtained with a finite-volume solver. These steps are discussed in more detail next and the interested reader can find more information about the automated optimization loop in Pimenta \textit{et al}.~\cite{Pimenta2020,Pimenta2018}

\subsection{\label{param}Geometry parametrization}

\begin{figure}[t]
\begin{center}
\includegraphics[scale=0.5]{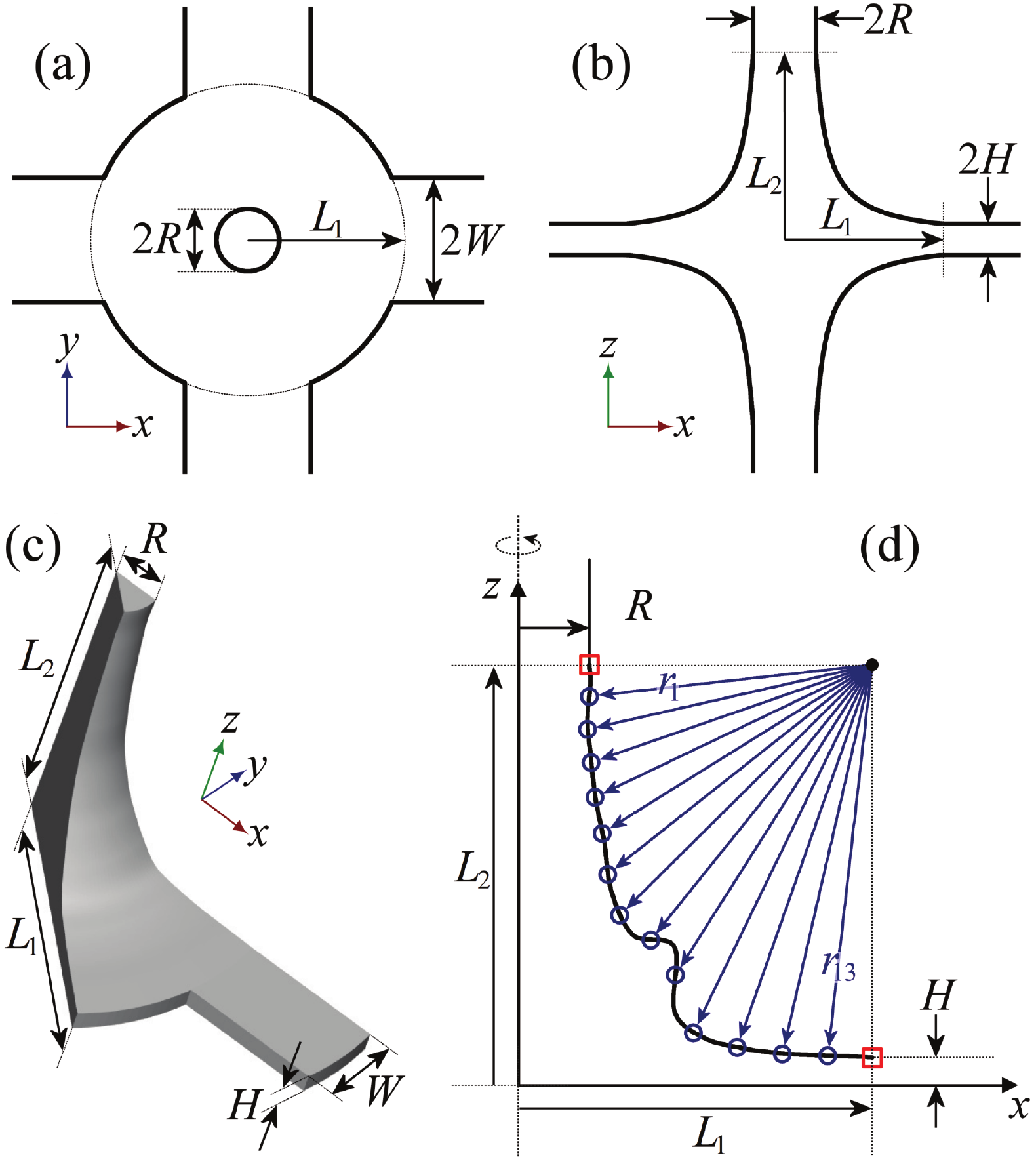}
\caption {Numerical scheme for the optimization of the 6-arm cross-slot. Initial guess of the geometry in (a) top-down view ($xy$ projection), (b) side view ($xz$ projection). The key geometric parameters are indicated where $R$ is the radius of the two circular channels, $W$ and $H$ are the half-width and half-height of the four planar channels, and $L_1$ and $L_2$ are the lengths of the optimized wall section along $x$ and $z$, respectively. (c) 3D rendering of 1/16$^\text{th}$ of the initial guess of the geometry, used for the numerical determination of the flow field.  (d) Schematic illustration of the optimization of the wall profile: the two red open squares are fixed in space at $(x,z) = (R,L_2)$ and $(x,z) = (L_1,H)$, whereas the thirteen open dark blue circles are free to move along the lines $r_1$ to $r_{13}$ radiating from the point  $(x,z) = (L_1,L_2)$ and form a Catmull-Rom interpolating spline that controls the shape of the wall.
} 
\label{numscheme}
\vspace{-0.2in}
\end{center}
\end{figure}

The shape optimization of a 6-arm cross-slot without any \textit{a priori} consideration is a difficult task due to the large number of design parameters that can arise. However, as discussed in Sec.~\ref{intro}, there is axial symmetry around the stretching direction in uniaxial extensional flows and around the compressive direction in biaxial extensional flows. Thus, in theory a geometry with axial symmetry is required to impose such flows. This would naturally result in two opposing circular inlets (outlets) over the compressive (extensional) $z$-axis in biaxial (uniaxial) extensional flows, that expand radially outward as they approach each other, to an outlet (inlet) over the $xy$ plane at $z=0$. However, any experimental realization of such a device requires both inlet and outlet connections for the fluid flow. As apparent in our first guess of the shape of an optimized 6-arm cross-slot geometry (Fig.~\ref{schematic}(d)), the two opposing circular inlets (outlets) centered on the $z$-axis are not problematic, however the four rectangular outlets (inlets) over the $xy$ plane inevitably break the axial symmetry of the geometry at some radial distance from the $z$-axis. This radial distance from the $z$-axis to the rectangular outlets (inlets) is parametrized in the optimization scheme as $L_1$, while the distance along $z$ from the $z=0$ plane to the circular inlets (outlets) is $L_2$ (see Fig.~\ref{numscheme}). The radius of the circular inlets (outlets) is $R$, while the half-width and half-height of the rectangular outlets (inlets) are $W$ and $H$, respectively. This approach results in a geometry with several planes of symmetry, which can be obtained through successive reflections of 1/16$^\text{th}$ of the whole geometry (Fig.~\ref{numscheme}(c)), which thus represents the elementary unit to be optimized. Note that the unit element selected for optimization could be optimized on its entire 3D surface, without any assumption of axial symmetry. However, this option is not undertaken because a large number of design parameters would still be necessary to fine-tune the shape of the wall. Instead, we prefer to keep axial symmetry in the main body of the elementary unit and simply connect it to the rectangular channel.

Under the approach described above, the wall of the elementary unit located between the circular channels and the rectangular channels is parameterized through $n=13$ movable points, which form a Catmull–Rom interpolating spline,~\cite{Catmull1974} as depicted in  Fig.~\ref{numscheme}(d). These points are evenly distributed through one quarter of an ellipse with center at $(x, z)=(L_1, L_2)$ and axes $(L_1-R)$ and $(L_2-H)$. The radii of these design points ($r_1$…$r_{13}$) can be adjusted by the optimizer to minimize the objective function, whereas their angle is kept fixed. The transition from the deformable section of the wall to the circular channels occurs at $z = L_2$, whereas the transition to the planar channels occurs at $x = L_1$. 

\subsection{\label{obj}Objective function}

In order to obtain an optimized 6-arm cross-slot device able to impose homogeneous uni- and biaxial extensional flows, it is necessary to translate the problem into a suitable mathematical formulation that can be handled by an optimizer. Although the definition of extensional flow holds in the entirety of 3D space, it is not feasible to aim at obtaining such a simple device as a 6-arm cross-slot imposing such kinematics in the whole of space. The most that can be done, and which has been done with other geometries,~\cite{Pimenta2018,Pimenta2020,Alves2008,Galindo2014,Zografos2016,Zografos2019} is to aim to achieve the desired flow over a limited region of space, and expecting uniformity in the neighborhood of that region. For the 6-arm cross-slot, we follow this approach by limiting the target region of extensional flow to the axes defined by each inlet/outlet. In the elementary unit simulated (Fig.~\ref{numscheme}(c)), we have two such inlets/outlets, one with a circular cross-section and another with a rectangular cross-section, and consequently two axes over which to impose the extensional flow. This can be done concretely by imposing the velocity profile that an extensional flow would have over those axes and defining an objective function to be minimized, which measures the difference between the actual velocity profiles and the theoretical ones:

\begin{equation}
\begin{split}
f_{obj}\lbrace r_{1...n}\rbrace_k & = \left( \frac{\alpha}{N}    \sum_{i=1}^N  \frac{\lvert {\bf{u}}_{i,k} - {\bf{u}}_{i,theo}\rvert}{\lvert {\bf{u}}_{i,theo}\rvert} \right)_x  \\ & + \left( \frac{\beta}{M}    \sum_{j=1}^M  \frac{\lvert {\bf{u}}_{j,k} - {\bf{u}}_{j,theo}\rvert}{\lvert {\bf{u}}_{j,theo}\rvert} \right)_z
.
\label{objfunc}
\end{split}
\end{equation}

Eq.~\ref{objfunc} corresponds to the formula adopted in this work to measure the objective function ($f_{obj}$) for each candidate geometry $k$, represented by its own array of design points $\lbrace r_{1...n}\rbrace_k$. Each of the two summations is relative to a given axis ($x$ and $z$), as the velocity profiles are different along each one. For each candidate geometry $k$, the velocity vector ${\bf{u}}_{i,k}$ (${\bf{u}}_{j,k}$), where ${\bf{u}} = (u,v,w)$, is sampled over $N$ ($M$) points and compared with the theoretical or expected velocity vector ${\bf{u}}_{i,theo}$ (${\bf{u}}_{j,theo}$) at the given position $i$ ($j$) of axis $x$ ($z$). The constants $\alpha$ and $\beta$ are used to weight the summation over each axis. Increasing the ratio of $\alpha / \beta$ or $\beta / \alpha$, increases the weighting of $f_{obj}$ over the $x$ or $z$ axis, respectively. In this work, we place equal importance on achieving homogeneous flows in both uniaxial and biaxial extension, so $\alpha$ and $\beta$ are simply set equal to 1. Varying $\alpha$ and $\beta$ in proportion has no effect on the final geometry but simply modifies $f_{obj}$ by the same factor. It should be noted that Eq.~\ref{objfunc} corresponds to a simple, but effective formulation of a bi-objective optimization, as the motion of a given design point can improve the velocity profile over one axis, but worsen the velocity profile over the other axis, i.e., there is a concurrent effect. More complex methods could be used in order to find the corresponding Pareto front (Nomad offers the possibility to do so),\cite{Digabel2011} but the simple single-objective formulation embodied by Eq.~\ref{objfunc} is able to provide good results. This is significantly different from previous works,~\cite{Pimenta2018,Pimenta2020,Alves2008,Galindo2014,Zografos2016,Zografos2019} where a single velocity profile was imposed, either because a single axis existed,~\cite{Pimenta2018,Pimenta2020,Zografos2016,Zografos2019} or because the velocity profiles over different axes were similar.~\cite{Alves2008,Galindo2014}   

For the 6-arm cross-slot, the theoretical velocity profiles are defined as:

\begin{equation}
\begin{split}
u_{i,theo} & = \frac{\dot\varepsilon x_i}{2} \text{ for }  x_i < L_1 \text{, } v_{i,theo}=w_{i,theo}=0 \\
w_{j,theo} & =-\text{min}\left( \dot\varepsilon z_j,2U \right) \text{, } u_{j,theo}=v_{j,theo}=0
,
\label{veltheo}
\end{split}
\end{equation}
and they correspond to a biaxial extensional flow with compression along the $z$-axis and extension along axes $x$ and $y$ (note that because biaxial and uniaxial flows are simply the reverse of each other kinematically, velocity profiles corresponding to uniaxial extensional flow could be used instead, without loss of generality). It should be noted that over $x$ a linear velocity profile is imposed up to $L_1$ but no constraint is imposed beyond that point. However, in practice the velocity profile transits to the constant velocity value imposed by the constant cross-section of the rectangular channels. The length over which that transition occurs is not constrained. On the other hand, we impose a sharp transition of velocity on the $z$-axis, which occurs at $z_j = 2U/\dot\varepsilon$, where $U$ is the mean flow velocity in the circular inlets and  $\dot\varepsilon$ is the extension rate. 

It can easily be shown that the average flow velocity across a cylindrical cross-section of increasing radius decreases with inverse proportionality to the radius. On the other hand, the rectangular channels impose a constant average and maximum velocity upon fixing $H$ and $W$. Therefore, $L_1$ needs to be carefully chosen for each pair $(H, W)$ in order to avoid severe constrictions of the geometry over the $x$-axis. In practice, $(H, W)$ are selected first in the ratio range $3 \leq W/H \leq 4$, then $L_1$ is adjusted such that the velocity profile in a geometry with constant height $H$ over $x$ lies above ${\bf{u}}_{x,theo}$ (this indicates that $H$ should be increased in order to locally decrease the average velocity). Taking the velocity value at $L_1$ in such geometry, the compression rate $\dot\varepsilon / 2$ is computed, which automatically defines the velocity profile over the $z$-axis, and hence the value of $L_2$.

\subsection{\label{CFD}CFD solution}

The velocity profiles that are compared against the theoretical profiles in the objective function (Eq.~\ref{objfunc}) are obtained after solving for the isothermal, incompressible flow of a Newtonian fluid in creeping flow conditions, which is governed by the continuity,

\begin{equation}
\nabla \cdot {\bf{u}}=0
,
\label{continuity}
\end{equation}
and momentum,
\begin{equation}
-\nabla p + \eta_s \nabla ^2 {\bf{u}} = 0
,
\label{momentum}
\end{equation}
equations where ${\bf{u}}$ is the velocity vector, $p$ is the pressure and $\eta_s$ is the constant viscosity of a Newtonian fluid. 

The governing equations are solved with the second-order finite-volume solver implemented in rheoTool,~\cite{Pimenta2017,Pimenta_rheoTool} which is based on OpenFOAM\textregistered. The geometry and mesh of the computational domain are built with the standard tools provided in OpenFOAM\textregistered. A validation study was carried out to ensure mesh independency of the results obtained and presented in this work.   

\begin{table}[!ht]
\caption{\label{Table1}Parameters used for the generation of various optimized 6-arm extensional flow geometries, and the final value of the minimized objective function in each case, $f_{obj,min}$. The wall profiles and predicted axial velocity profiles for each geometry are shown in Fig.~\ref{geometries}.  }
\begin{ruledtabular}
\begin{tabular}{c c c c c c}
Geometry   &   $L_1/R$  &   $L_2/R$     &   $W/R$  &  $H/R$   & $f_{obj,min}$ \\
\hline
A & 5 & 6.5 & 1.5 & 0.5  & 0.0276         \\
B & 5.5 & 8 & 1.75 & 0.5  & 0.0307   \\
C &  5 & 6 & 1.6 & 0.4  & 0.0238 \\
D & 5.5 & 9 & 2 & 0.5  & 0.0263  \\

\end{tabular}
\end{ruledtabular}
\end{table}

\begin{figure*}[!ht]
\begin{center}
\includegraphics[scale=0.53]{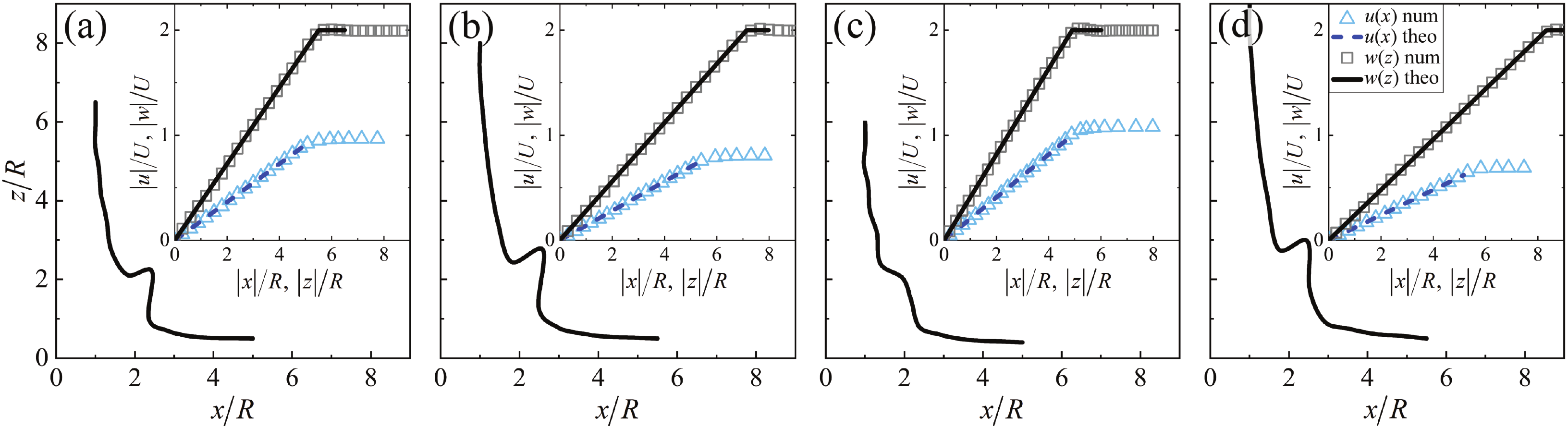}
\caption {Wall profiles between the points $(x,z)=(R,L_2)$ and $(x,z)=(L_1,H)$ for the four optimizations performed under Newtonian creeping flow conditions and with the values of $L_1$, $L_2$, $W$ and $H$ given in Table~\ref{Table1}. Figure parts (a) to (d) correspond to Geometries A to D in Table~\ref{Table1}, respectively. The respective inset plots show the corresponding numerically-predicted streamwise velocity profiles along the flow axes (data points), compared with the ideal theoretical profiles (solid lines).
} 
\label{geometries}
\end{center}
\end{figure*}

\subsection{\label{geo}Optimized geometries}

In this work, optimizations have been performed according to the scheme previously described using several combinations of the design parameters $L_1$, $L_2$, $H$, and $W$ (see Table~\ref{Table1}), with each combination of parameters leading to a distinct geometry. In Fig.~\ref{geometries}, the wall profile resulting from each imposed set of design parameters is shown, along with the corresponding axial velocity profiles obtained from the numerically-solved flow field. In each case, there is an excellent agreement between the obtained velocity profile and the theoretical target. 

\subsection{\label{Viscoelastic}Viscoelastic flow simulations}

The four optimizations of the 6-arm cross-slot all yield geometries with similar performance, as evident from the similar respective values of the minimized objective function $f_{obj,min}$ given in Table~\ref{Table1}, and the close match between the numerical and theoretical axial velocity profiles in each case (Fig.~\ref{geometries}).  Hence, we only select one of them for fabrication and experimental verification. The most obvious suitable candidate geometry for the fabrication is Geometry C, since it has the lowest value of $f_{obj,min}$ and it also does not possess the non-monotonicity in the wall profile that occur in Geometries A, B and D at around $x/R=2$ (Fig.~\ref{geometries}). Such deep concavities would not only be difficult to reproduce accurately by our fabrication method (see Sec.~\ref{Fab}), but could easily trap air bubbles during fluid loading, affecting the resulting flow field during experimentation. 

Since the ultimate intented application of the geometry is focused on extensional rheometry of viscoelastic fluids, it is important to assess the impact of the fluid rheology on either the form of the geometry (or on the flow field imposed by the geometry). Note that, in principle the geometry could be optimized using a viscoelastic rather than a Newtonian flow, as was performed in 2D for the planar OSCER geometry (in that case showing almost negligible differences in the resulting shape or velocity profiles).~\cite{Alves2008} However, due to the large number of iterations ($\sim 100$'s) needed to minimize the objective function (Eq.~\ref{objfunc}), this approach is prohibitively computationally costly in 3D. Instead, we opt to demonstrate that our selected geometry (as optimized based on a Newtonian flow) also imposes essentially the same flow field regardless of the rheology of the fluid. This is an important requirement to ensure the device will be suitable for the characterization of different types of fluids. For this reason, we perform numerical simulations in Geometry C using the Oldroyd-B and Phan-Thien and Tanner viscoelastic constitutive models, as described next.

\subsubsection{Governing equations}

The non-Newtonian flow is described by the incompressible and isothermal Cauchy equations coupled with a constitutive equation, which accounts for the contribution of the non-Newtonian stresses. Neglecting inertia, the continuity equation is given above (Eq.~\ref{continuity}), while the momentum equation becomes:  

%\begin{equation}
%\nabla \cdot \textnormal{\bf{u}}=0
%,
%\label{continuity}
%\end{equation}

\begin{equation}
\nabla \cdot (-p\textnormal{\bf{I}}+\pmb{\uptau}+\eta_s \pmb{\dot\upgamma})=0
,
\label{non-Newt_momentum}
\end{equation}
where, $\textnormal{\bf{I}}$ is the identity tensor, and $\pmb{\uptau}$ is the non-Newtonian contribution to the total stress tensor.

The  constitutive equation for a Phan-Thien and Tanner (PTT) fluid is expressed as:
\begin{equation}
\lambda \biggl[\frac{\partial{\pmb{\uptau}}}{\partial t} + {\bf{u}}\cdot \nabla{\pmb{\uptau}} -(\nabla{\bf{u}})^T\cdot{\pmb{\uptau}} - {\pmb{\uptau}} \cdot \nabla{\bf{u}} \biggr]+Y{\pmb{\uptau}}=\eta_p{\pmb{\dot\upgamma}}
,
\label{constitutive}
\end{equation}
 where $\lambda$ is the relaxation time, and $\eta_p$ is the polymeric viscosity coefficient. The deformation rate tensor, $\pmb{\dot\upgamma}$ ($=2\textnormal{\bf{D}}$), is defined as: 

\begin{equation}
\pmb{\dot\upgamma}=\nabla \textnormal{\bf{u}} + (\nabla \textnormal{\bf{u}})^T
,
\label{deformationrate}
\end{equation}
where the superscript ``$T$'' denotes the transpose operator. The function $Y$ is given as: 

\begin{equation}
Y= 1 + \varepsilon \frac{\lambda \textnormal{tr}(\pmb{\uptau})}{\eta_p}
,
\label{Y_function}
\end{equation}
where $\textnormal{tr}(\pmb{\uptau})$ denotes the trace of $\pmb{\uptau}$, and $\varepsilon$ is a parameter that governs the rheological response of the fluid and will be discussed below.

The usual no-slip and no-penetration boundary conditions (i.e., $\textnormal{\bf{u}}={\bf{0}}$) are imposed on all surfaces of the channel. At the channel inlets, we impose fully-developed velocity and stress fields. At the channel outflows, we apply the open boundary condition (OBC). \cite{Papanastasiou1992} Finally, we apply the usual symmetry conditions at all symmetry planes.

\subsubsection{Oldroyd-B model}
\label{OB}

The Oldroyd-B (O-B) model is retrieved when setting $\varepsilon = 0$ in Eq.~\ref{Y_function}, leading to $Y=1$. Under steady simple shear, the O-B model predicts a constant viscosity, $\eta_0 = \eta_p + \eta_s$, while the solvent-to-total viscosity ratio is defined as $\beta =\eta_s/\eta_0$. Under steady extension (uniaxial, biaxial, or planar) the model predicts an extensional viscosity that is almost constant at low values of the strain rate, but which tends towards infinity as the dimensionless strain rate, or Weissenberg number, $\text{Wi}~=~\lambda\dot\varepsilon\rightarrow0.5$ ($\text{Wi}= \lambda\dot\varepsilon_B\rightarrow0.5$ in biaxial extension). We test two typically used values of $\beta$ ($1/9$ and 0.59), at two values of the Weissenberg number ($\text{Wi}=0.2$ and 0.4).

\subsubsection{Linear Phan-Thien and Tanner model}
\label{l-PTT}

The linear version of the simplified PTT model (l-PTT) \cite{Phan1977} is retrieved from Eq.~\ref{Y_function} when $\varepsilon > 0$. The l-PTT model predicts shear-thinning effects in steady simple shear, and a bounded extensional viscosity in steady extension (uniaxial, biaxial, or planar). With increasing $\varepsilon$, the fluid becomes less strain-hardening and the onset of shear thinning is translated to lower values of the shear rate. We test typically used values of $\beta = 1/9$ and $\varepsilon = 0.02$, at two values of the Weissenberg number ($\text{Wi}=0.4$ and 0.8).

\begin{figure*}[!ht]
\begin{center}
\includegraphics[scale=0.7]{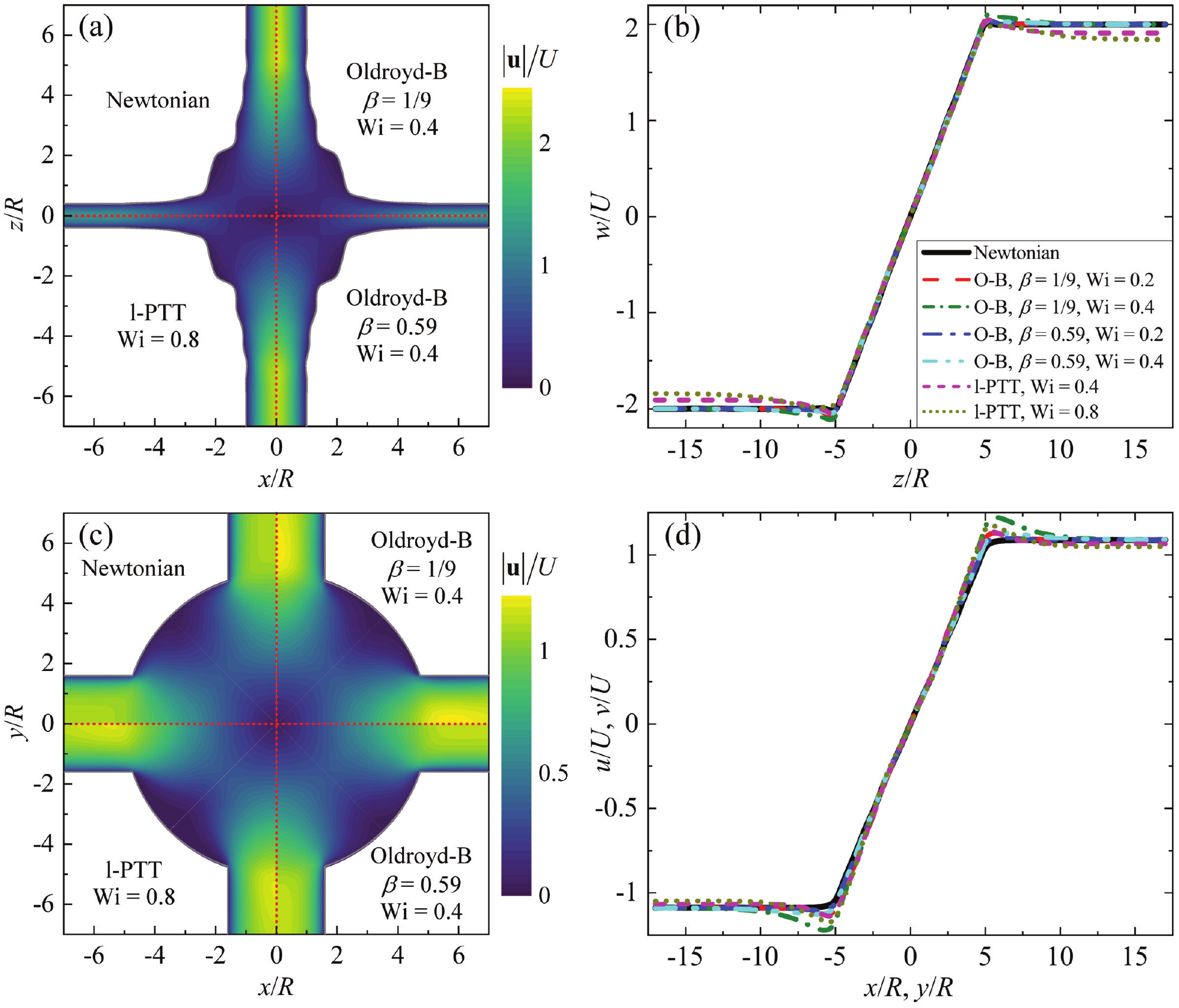}
\vspace{-0.1in}
\caption {Results of viscoelastic flow simulations with the Oldroyd-B (O-B) and linear Phan-Thien and Tanner (l-PTT) constitutive models in Geometry C, compared with the creeping flow Newtonian prediction. (a) Normalized velocity magnitude field ${\abs{\bf{u}}}/U$ in uniaxial extension, where each quadrant shows the prediction of the fluid model indicated. (b) Normalized streamwise velocity profiles predicted for each of the examined fluid models along the extensional ($z$) axis in uniaxial extension. (c) Normalized velocity magnitude field in biaxial extension, where each quadrant shows the prediction of the fluid model indicated. (d) Normalized streamwise velocity profiles predicted for each of the examined fluid models along the extensional ($x$, or $y$) axes in biaxial extension.
} 

\label{viscoelastic}
\end{center}
\end{figure*}

\subsubsection{Numerical Method}

The Petrov-Galerkin stabilized Finite Element Method for Viscoelastic flows (PEGAFEM-V) \cite{Varchanis2019, varchanis2020pegafem} is used to solve the governing equations. We solve directly for the steady state solution, neglecting the time derivative in Eq.~\ref{constitutive}. The flow variables, $\textnormal{\bf{u}}$, $p$, and $\pmb{\uptau}$, are interpolated by linear tetrahedra in a structured mesh. A validation study was again carried out to ensure mesh independency of the results at the values of $\text{Wi}$ examined herein (see Appendix~\ref{appendix}).

In Fig.~\ref{viscoelastic}, we present the results of the viscoelastic flow simulations performed in Geometry C. The upper left quadrant of Fig.~\ref{viscoelastic}(a) shows normalized velocity magnitude fields in the $y=0$ plane predicted for creeping Newtonian flow in uniaxial extension, while the remaining quandrants of the figure show the predictions of the viscoelastic fluid models at the highest values of the Weissenberg numbers tested. It is clear that the flow field predicted by the viscoelastic fluid models does not deviate significantly from the Newtonian prediction. Indeed, over the optimized region of the geometry, profiles of the streamwise velocity along the extensional axis ($w(z)$), show excellent agreement with the Newtonian prediction for both the O-B and and the l-PTT models at all values of $\text{Wi}$ (Fig.~\ref{viscoelastic}(b)). Compared with the Newtonian velocity profile, we can observe only a slight overshoot in the velocity for the viscoelastic models near $z=\pm 5R$, and a slight reduction of the fully-developed centerline flow velocity for the l-PTT model within the circular outlet channels for $\abs{z} > 5R$ (which is due to the shear thinning). 

Fig.~\ref{viscoelastic}(c) shows normalized velocity magnitude fields in the $z=0$ plane predicted for flow in biaxial extension, again divided into quandrants depicting creeping Newtonian flow (upper left) and the predictions of the viscoelastic fluid models at the highest values of the Weissenberg numbers tested. Again there is visibly rather close agreement between the Newtonian and the viscoelastic predictions of the velocity field, although it is noticeable that the velocity magnitude tends to be slightly higher for the viscoelastic models close to the entrances of the planar outlet channels. Profiles of the streamwise flow velocity along the outlet axes in biaxial extension (i.e., $u(x) (\equiv v(y))$) again demonstrate rather close agreement between the predictions for Newtonian and viscoelastic flows (Fig.~\ref{viscoelastic}(d)). The viscoelastic flow predictions deviate slightly from the Newtonian prediction near the limits of the optimized region (i.e., $x=y=\pm 5R$) and a small overshoot is evident immediately inside the planar outlet channels. Also, we can observe that for the l-PTT fluid the fully-developed centerline flow velocity within the outlet channels is slightly lower than for the Newtonian case (due to the shear thinning). 

In general, even employing rather stringent viscoelastic models that give an unbounded response to extensional flow (i.e., O-B) and that account for the combination of elastic effects and a shear thinning viscosity (i.e., l-PTT), the flow field imposed by Geometry C does not deviate significantly from that under Newtonian creeping flow. The geometry provides the desired uniaxial and biaxial extensional flow fields given a variety of rheological conditions and imposed Weissenberg numbers up to $\text{Wi} = 0.8$. Therefore it appears to be a promising candidate geometry for an extensional rheometer based on uniaxial and biaxial elongation.

\begin{figure*}[!ht]
\begin{center}
\includegraphics[scale=0.5]{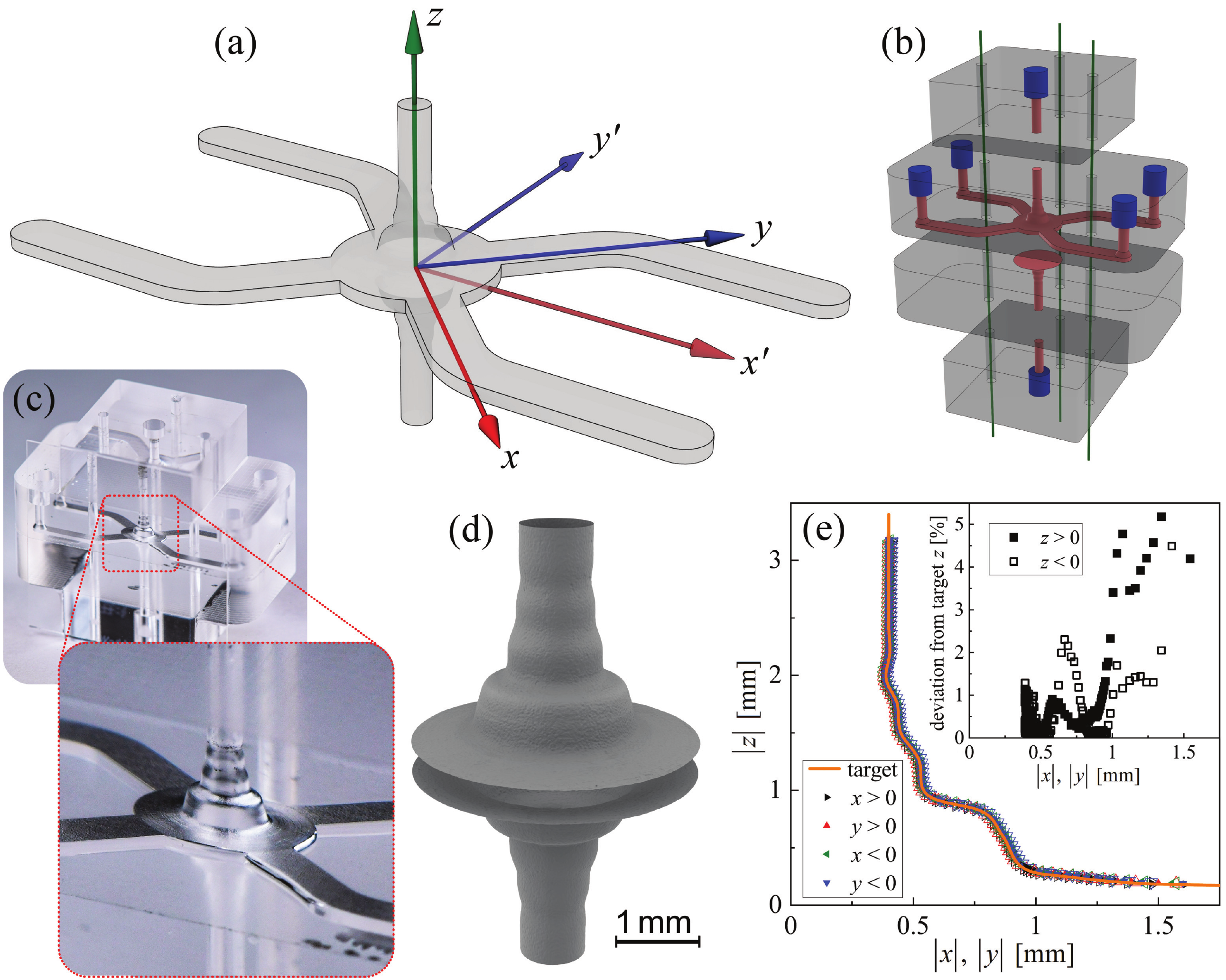}
\caption {Experimental realization of an optimized 6-arm cross-slot based on Geometry C (see Table~\ref{Table1} and Fig.~\ref{geometries}). (a) 3D rendering of the design to which the geometry is fabricated, also showing the standard coordinate system $(x,y,z)$ and the $45^{\circ}$ rotated coordinate system $(x',y',z)$ used in the experiments (see main text). (b) Exploded schematic view of the four glass parts constituting the microfluidic device. For clarity, the main flow channel is colored in red and the inlet and outlet ports are colored in blue. The four parts are assembled on locating pins, indicated by the vertical (green) lines. (c) Photograph of the actual assembled device, with a zoomed-in view of the central portion. (d) 3D rendered image constructed from the output of an X-ray $\upmu$-CT scan of the device. (e) Comparison between wall profiles obtained from the $\upmu$-CT scan and the target design shown in part (a). Insert shows the percentage deviation from the target $z$ value as a function of the $x$ or $y$ coordinate.
} 
\label{fab}

\end{center}
\end{figure*}

\section{Experimental Methods}
\label{ExpMeth}

\subsection{Microfluidic uni- and biaxial extensional flow device}
\label{Fab}

In the remainder of the paper, we will focus on the experimental realization of Geometry C and the verification of its performance based on Newtonian fluid flow. The fabrication of the device, which from now on we will refer to as the OUBER (Optimized-shape Uni- and Biaxial Extensional Rheometer), is achieved by the technique of selective laser-induced etching (SLE) in fused silica glass. \cite{Gottmann2012,Meineke2016,Burshtein2019} SLE is a two-step subtractive 3D printing technique for use with transparent substrates (typically glass). SLE in fused silica enables the fabrication of arbitrarily shaped microchannels with high resolution ($\sim \mathcal{O}(1~\upmu$m)) in a rigid high modulus substrate with excellent optical clarity. In brief, the process involves the use of a scanning femtosecond laser to irradiate the volume to be removed (i.e., the internal volume of the microchannel) from a block of pristine fused silica substrate. The laser irradiation is performed using a commercially available LightFab 3D printer (LightFab GmbH). Subsequent to the laser scanning, the fused silica block is ultrasonicated in potassium hydroxide at $80^{\circ}$C, and the irradiated material is selectively removed.

A 3D rendering of the design of the OUBER geometry, used to define the volume scanned by the femtosecond laser, is shown schematically in Fig.~\ref{fab}(a) (minus the inlet and outlet ports). The geometry is scaled such that the circular channels have a radius $R=0.4$~mm. Thus the half-width and half-height of the four planar channels are $W=0.64$~mm and $H=0.16$~mm, respectively (Table~\ref{Table1}). The geometry is divided into four parts along $z$, each of which has a thickness $\delta z = 5$~mm (the maximum thickness of substrate that can be used in the LightFab instrument). The reason for this division is because the laser scanning has a higher resolution in $x$ and $y$ than in $z$. Therefore, circular holes are formed with the highest resolution in $xy$ planes. Scaling down the device dimensions to fit the entire channel within a single 5~mm thick substrate is not currently practical. The four individual pieces are assembled on locating pins and bonded together using ultra-violet-curing epoxy resin (see exploded view of the assembly in Fig.~\ref{fab}(b)). A photograph of the fully-assembled glass device is provided in Fig.~\ref{fab}(c). X-ray microtomography ($\upmu$-CT) scanning of the central cross-over region of the channel is performed using a Zeiss Xradia 510 Versa 3D X-ray microscope operated with the Zeiss Scout-and-Scan Control System software. The $\upmu$-CT scan data is reconstructed using Amira analysis software (Thermo Fisher) and exported to Rhinoceros 3D modeling software (Robert McNeel and Associates) to construct the 3D rendered image in Fig.~\ref{fab}(d). Eight surface profiles extracted from the $\upmu$-CT scan of the channel at four azimuthal angles (corresponding to the positive and negative $x$, $y$ and $z$ directions) are compared to the target channel profile in Fig.~\ref{fab}(e). The inset to Fig.~\ref{fab}(e) shows the root-mean-square deviation $s_z$ of the extracted profiles from the target (expressed as a percentage of the local target $z$), demonstrating the excellent fidelity of the fabrication to the design ($s_z \lesssim 0.05 z ~\forall x,y$).

Note that the natural choice of coordinate system has the $x$ and $y$ axes aligned with adjacent planar channels and the $z$ axis aligned with the circular channels (as was done during the optimization step, c.f., Fig.~\ref{numscheme}). However, since the axisymmetry of the device is broken by the four planar inlet/outlet channels, in the following we will also consider a coordinate system rotated by $45^{\circ}$ about the $z$-axis (as shown in Fig.~\ref{fab}(a)), i.e., such that $x' = \frac{1}{\sqrt{2}}(x+y)$, $y' = \frac{1}{\sqrt{2}}(y-x)$. Comparison of the flow profiles along the standard and the $45^{\circ}$ rotated axes will reveal the extent to which the flow field remains axisymmetric.

\subsection{Test fluid}

Due to the surface curvature of the 3D OUBER device, see Fig.~\ref{fab}, clear imaging inside of the device (e.g., for performing flow velocimetry, as described below) requires that the channel be filled with a fluid of similar refractive index $RI$ as the fused silica glass. A sufficiently good match is achieved with a mixture 89.6~wt\% glycerol and 10.4~wt\% water, with $RI=1.4582$ at $25^{\circ}$C  (measured using an Anton-Paar Abbemat MW refractometer operating at 589~nm). This is close to the value of $RI=1.4584$ expected for fused silica under the same conditions.~\cite{Malitson1965} The 89.6:10.4~wt\% glycerol:water mixture has density $\rho = 1231$~kg~m$^{-3}$ and viscosity $\eta = 0.143$~Pa~s.

\subsection{Flow control}
\label{control}

Flow is driven through the microfluidic OUBER device by using 29:1 gear ratio neMESYS low pressure syringe pumps (Cetoni, GmbH) to control the volumetric flow rate through each individual inlet/outlet channel. For uniaxial (biaxial) extensional flow, two pumps are used to impose a volumetric flow rate $Q$ through the two circular outlet (inlet) channels, while four pumps impose a volumetric flow rate $Q/2$ through the four inlet (outlet) channels. The pumps are fitted with Hamilton Gastight syringes of appropriate volumes so that the specified ``pulsation free'' dosing rate of each pump is always exceeded. Connections between the syringes and the microfluidic device are made using flexible Tygon tubing. 

We consider the characteristic average flow velocity in the OUBER device as that in the circular channels, $U=Q/\uppi R^2$. The Reynolds number of the flow is defined as $\text{Re}=
2 \rho U R / \eta$, and the maximum value reached in the experiments is $\text{Re}~\approx~0.1$. Since $\text{Re}<1$ in all of the experiments, inertial effects in the flow are considered irrelevant. The expected extensional rates obtained from the numerical flow velocity profiles given in Fig.~\ref{geometries}(c) are $\dot\varepsilon=0.4U/R$ and $\dot\varepsilon_B=0.2U/R$ in uniaxial and biaxial extension, respectively.

\subsection{Microtomographic particle image velocimetry}
\label{PIV}

The flow field in the vicinity of the stagnation point of the OUBER device is measured volumetrically using microtomographic particle image velocimetry ($\upmu$-TPIV).~\cite{Carlson2021} Measurements are conducted using a LaVision FlowMaster system (LaVision GmbH), comprised of a stereomicroscope (SteREO V20, Zeiss AG, Germany) with dual high speed cameras (Phantom VEO 410, 1280 x 800 pixels) imaging a fluid volume illuminated by a coaxial Nd:YLF laser (dual-pulsed, 527~nm wavelength). The fluid is seeded with $3.2~\upmu$m diameter fluorescent particles (Fluoro-Max, Thermo Scientific), with excitation/emission wavelength 542/612~nm, to a visual concentration of $\approx 0.04$ particles-per-pixel. 

\begin{figure*}[!ht]
\begin{center}
\includegraphics[scale=0.8]{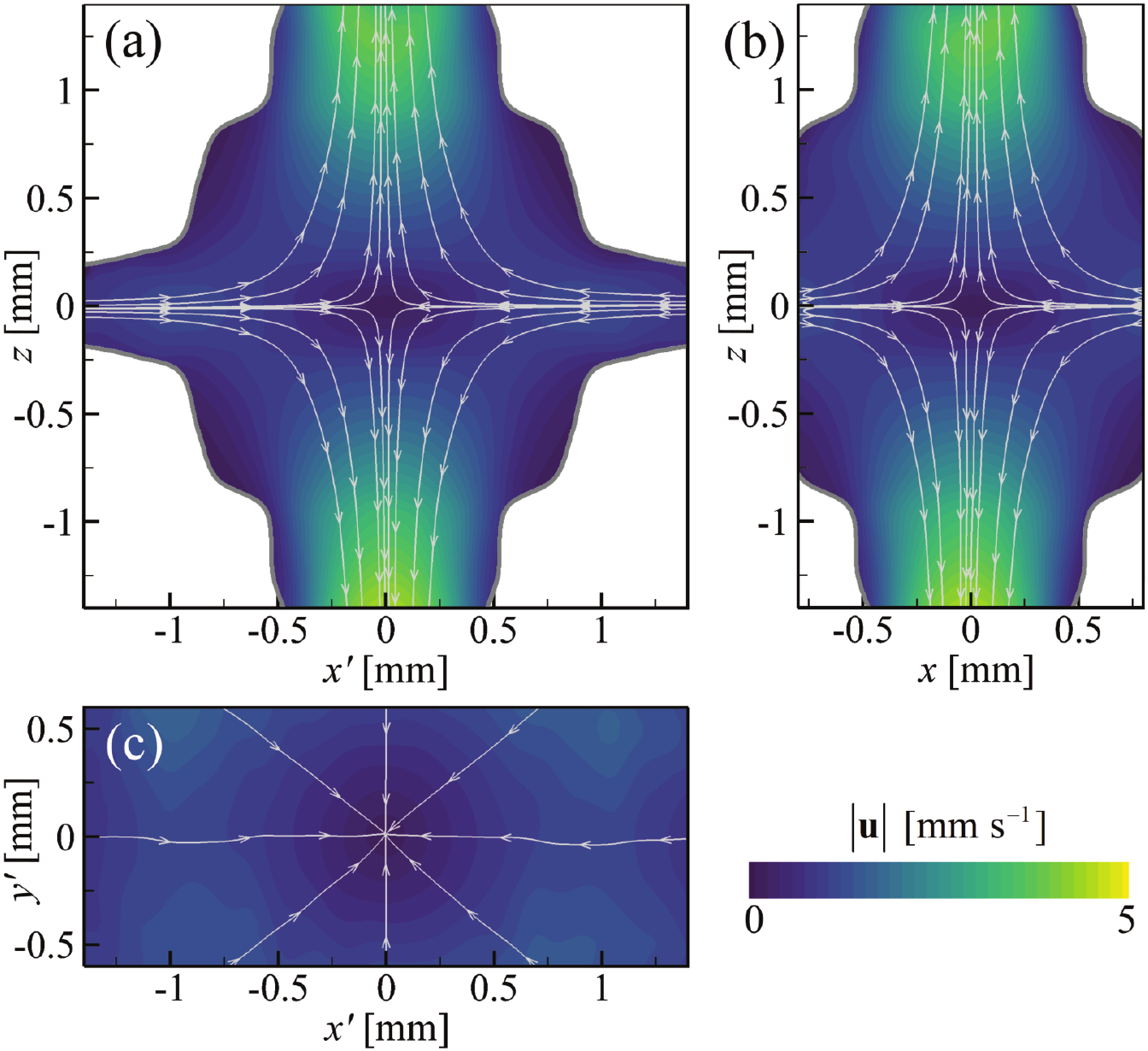}

\caption {Experimental flow field measured using $\upmu$-TPIV for uniaxial extension of a Newtonian fluid at $Q = 0.1$~mL~min$^{-1}$ ($\text{Re}\approx0.02$) in the OUBER device. Velocity magnitude fields with superimposed streamlines in (a) the $y'=0$ plane, (b) the $y=0$ plane, and (c) the $z=0$ plane. 
} 
\label{UniPIV}
\end{center}
\end{figure*}
\begin{figure*}[!ht]
\begin{center}
\includegraphics[scale=0.5]{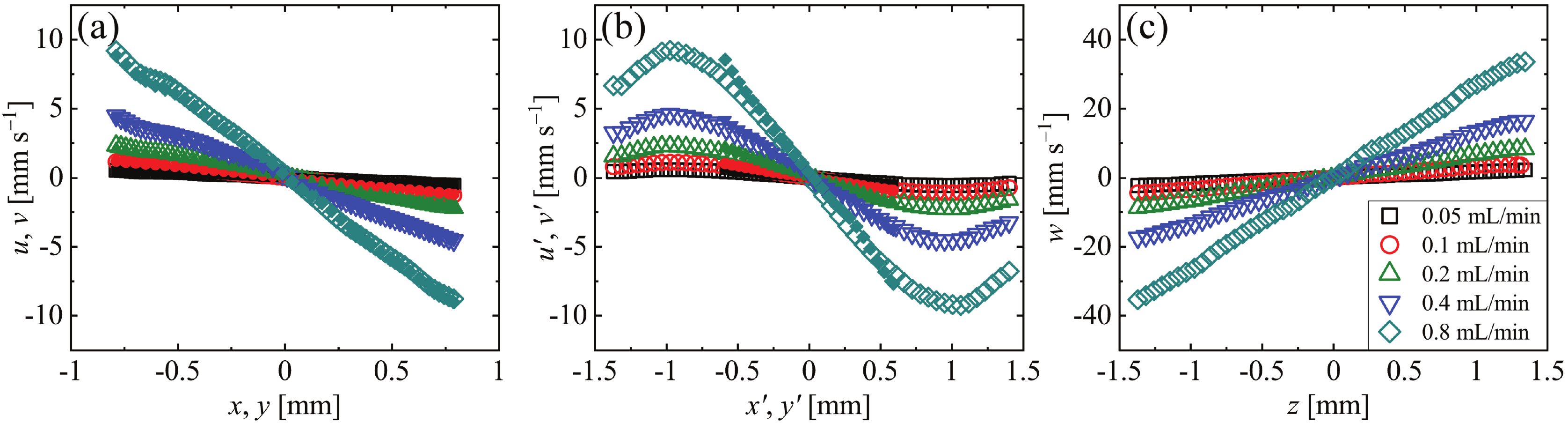}

\caption {Experimental streamwise axial velocity profiles extracted from $\upmu$-TPIV measurements for uniaxial extension of a Newtonian fluid at various volumetric flow rates $Q$ in the OUBER device. (a) $u(x)\rvert_{y=z=0}$ (open symbols) and $v(y)\rvert_{x=z=0}$ (closed symbols), (b) $u'(x')\rvert_{y'=z=0}$ (open symbols) and $v'(y')\rvert_{x'=z=0}$ (closed symbols), and (c) $w(z)\rvert_{x=y=0}$. 
} 
\label{UniProfiles}
\end{center}

\end{figure*}

Optical access to the stagnation point of the OUBER device is possible along the $y'$ direction, between two of the planar channels (see Fig.~\ref{fab}). We focus on the $y'=0$ plane of the OUBER device at $30\times$ magnification, which enables reliable recording of the flow in a rectangular cuboidal volume defined by the limits $-1.4\leq x' \leq 1.4~\text{mm}$, $-0.6\leq y' \leq 0.6~\text{mm}$, $-1.4\leq z \leq 1.4~\text{mm}$. The flow is recorded as single-frame images captured at a rate that is varied inversely to the imposed flow rate such that no particle moves more than 8 pixels between consecutive frames. Images are pre-processed with local background subtraction and Gaussian smoothing at $3 \times 3$~pixels. 3D calibration is performed by capturing reference images of a micro-grid at the planes $y' = \pm 1000~\upmu$m and $y'= 0~\upmu$m, and a coordinate system is interpolated between these planes using a third-order polynomial. Particle positions in 3D are reconstructed from the images using four iterations of the Fast MART (Multiplicative Algebraic Reconstruction Technique) algorithm,~\cite{Worth2008, Atkinson2009} followed by iterations of Sequential MART (SMART),~\cite{Atkinson2009} implemented in the commercial PIV software (DaVis 10.1.2, Lavision GmbH). We conclude the algorithm with five iterations of the Sequential Motion Tracking Enhancement (SMTE) method \cite{Novara2010,Lynch2015} to reduce the incidence of spurious ``ghost'' particles that arise due to randomly overlapping lines of sight, \cite{Elsinga2006} and which thus do not correlate in time. Volume self-calibration \cite{Wieneke2008} is employed to improve the accuracy of reconstruction. Particle displacements between particle volumes are obtained using a multi-grid iterative cross-correlation technique, with the final pass at $32 \times 32 \times 32$ voxels with 75\% overlap yielding velocity vectors  $\bf{u}$ on a cubic grid of $33.2~\upmu$m spacing. The obtained components of $\bf{u}$ are labeled as $u'$, $v'$, and $w$, in the $x'$, $y'$, and $z$ directions, respectively. Since the measured flows are time-steady, to reduce measurement noise typically 50 vector fields are averaged (note that ghost particle intensity is converged after averaging of $\approx 5$ frames). In one particular case (for an imposed volumetric flow rate $Q=0.1$~mL~min$^{-1}$), 200 vector fields are averaged in order to obtain sufficiently smooth data for computation of derived quantities. Subsequent to data acquisition, the software Tecplot 360 (Tecplot Inc., WA) is used for generation of contour plots, streamline traces, computation of the vector components $u$ and $v$ (in the respective $x$ and $y$ directions), and for extraction of velocity profiles, etc.

\begin{figure*}[!ht]
\begin{center}
\includegraphics[scale=0.8]{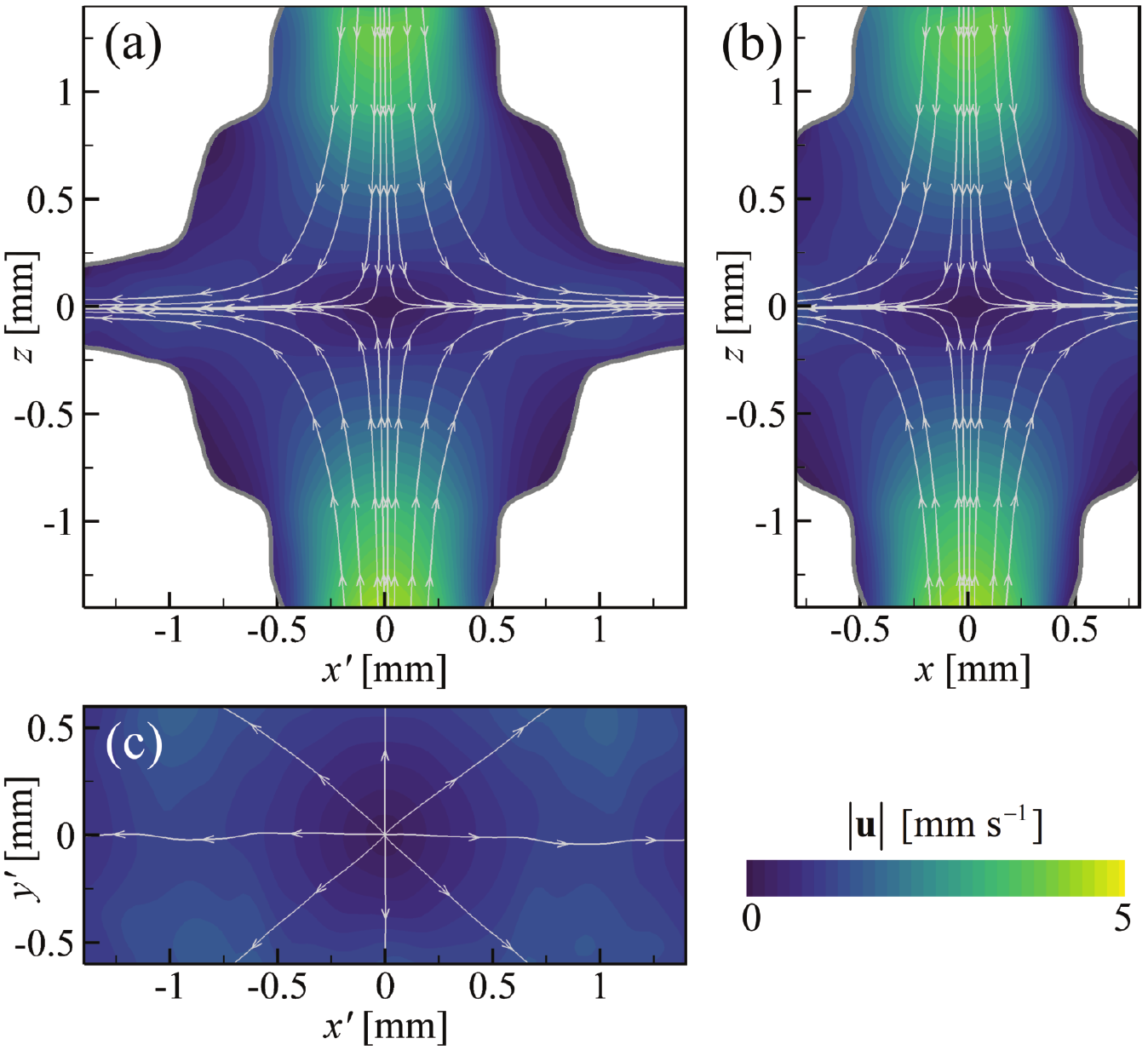}

\caption {Experimental flow field measured using $\upmu$-TPIV for biaxial extension of a Newtonian fluid at $Q = 0.1$~mL~min$^{-1}$ ($\text{Re}\approx0.02$) in the OUBER device. Velocity magnitude fields with superimposed streamlines in (a) the $y'=0$ plane, (b) the $y=0$ plane, and (c) the $z=0$ plane. 
} 
\label{BiPIV}
\end{center}
\end{figure*}
\begin{figure*}[!ht]
\begin{center}
\includegraphics[scale=0.5]{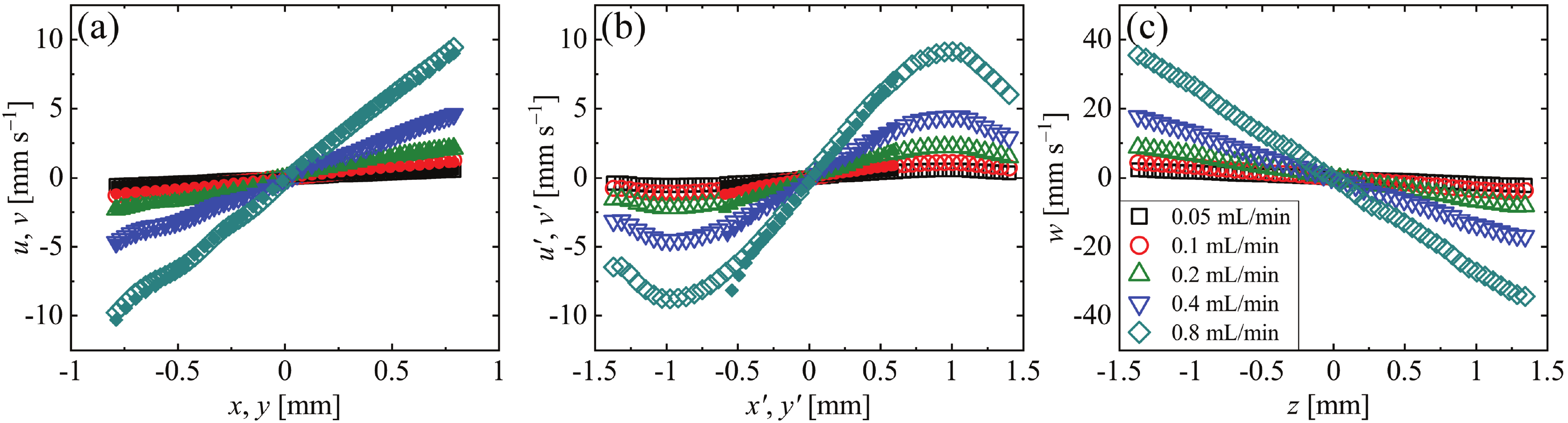}

\caption {Experimental streamwise axial velocity profiles extracted from $\upmu$-TPIV measurements for biaxial extension of a Newtonian fluid at various volumetric flow rates $Q$ in the OUBER device. (a) $u(x)\rvert_{y=z=0}$ (open symbols) and $v(y)\rvert_{x=z=0}$ (closed symbols), (b) $u'(x')\rvert_{y'=z=0}$ (open symbols) and $v'(y')\rvert_{x'=z=0}$ (solid symbols), and (c) $w(z)\rvert_{x=y=0}$. 
} 
\label{BiProfiles}
\end{center}
\end{figure*}

\section{Experimental Results}
\label{Res}

\subsection{Newtonian flow field characterization}

In Fig.~\ref{UniPIV} we present experimental velocity magnitude fields (${\abs{\bf{u}}}=\sqrt{u'^2 + v'^2 + w^2}$) with superimposed projected streamlines for uniaxial extensional flow in the OUBER device at an imposed volumetric flow rate of $Q=0.1$~mL~min$^{-1}$ (which corresponds to $\text{Re}\approx0.02$). The flow field as seen in the $y'=0$ plane is shown in Fig.~\ref{UniPIV}(a). The flow velocity decreases towards zero as the projected streamlines approach the stagnation point along the $x'$ direction and increases as streamlines separate from each along the $z$ direction. Within the available field of view in the $y=0$ plane (Fig.~\ref{UniPIV}(b)), the flow field appears similar to that in Fig.~\ref{UniPIV}(a), as would be expected if the flow were ideally axisymmetric. In the $z=0$ plane (Fig.~\ref{UniPIV}(c)) the field of view is quite restricted looking into the geometry along the $y'$ direction, however, within the accessible field of view we observe approximately circular contours of $\abs{\bf{u}}$ and streamlines that approach each other radially and converge at the stagnation point at $x'=y'=0$.

Profiles of the streamwise axial velocity components $u(x)\rvert_{y=z=0}$ and $v(y)\rvert_{x=z=0}$ are shown by the open and closed symbols (respectively) for several imposed values of $Q$ in Fig.~\ref{UniProfiles}(a). Clearly (over the accessible field of view in $x$ and $y$), the velocity profiles along the two orthogonal inlet axes are similar, with a nearly constant slope that  becomes steeper as $Q$ is increased. Along the $x'$ and $y'$ axes (Fig.~\ref{UniProfiles}(b)), we have a different field of view in each direction. However, the axial profiles of the streamwise velocity components $u'(x')\rvert_{y'=z=0}$ (open symbols) and $v'(y')\rvert_{x'=z=0}$ (closed symbols) appear to be similar and to have an approximately constant slope as far as can be measured along $y'$ (i.e., for $\abs{x'} = \abs{y'} \leq 0.6~\text{mm}$). With increasing distance from the $z$-axis beyond $\pm 0.6~\text{mm}$, the profiles of $u'(x')$ (open symbols) pass through local extrema before decreasing in magnitude. This is because the flow along the $x'$ (and also the $y'$) axis is directed towards the boundary of the flow cell located at  $L_1 = 5R (=2~\text{mm})$, where the flow velocity must vanish. Fig.~\ref{UniProfiles}(c) shows profiles of the streamwise velocity component along the outlet axis $w(z)\rvert_{x=y=0}$. Over the measurable range of $z$, the outlet axis velocity profiles are linear, with a slope that increases in proportion with $Q$, as expected. Note that $u$, $v$ and $w$ all vanish at $(x,y,z)=(0,0,0)$, which is the expected location of the stagnation point.

In Fig.~\ref{BiPIV}, we present experimental velocity magnitude fields with superimposed projected streamlines for biaxial extensional flow in the OUBER device, here again at an imposed volumetric flow rate of $Q=0.1$~mL~min$^{-1}$  (or $\text{Re}~\approx~0.02$). Fig.~\ref{BiPIV}(a), Fig.~\ref{BiPIV}(b), and Fig.~\ref{BiPIV}(c) illustrate the biaxial extensional flow field as observed in the $y'=0$ plane, the $y=0$ plane, and the $z=0$ plane, respectively. Comparison with Fig.~\ref{UniPIV}, for uniaxial extension at the same imposed $Q$, shows that the velocity magnitude fields are almost identical, however the direction of the streamlines is reversed. This is as expected given the kinematic reversibility of uniaxial and biaxial extension. Accordingly, profiles of the streamwise axial velocity components in biaxial extension $u(x)\rvert_{y=z=0}$ and $v(y)\rvert_{x=z=0}$ (Fig.~\ref{BiProfiles}(a)),  $u'(x')\rvert_{y'=z=0}$ and $v'(y')\rvert_{x'=z=0}$ (Fig.~\ref{BiProfiles}(b)), and $w(z)\rvert_{x=y=0}$ (Fig.~\ref{BiProfiles}(c)), are essentially just mirror images of those obtained in uniaxial extension (Fig.~\ref{UniProfiles}(a,b,c)).

\begin{figure}[!ht]
\begin{center}
\includegraphics[scale=0.7]{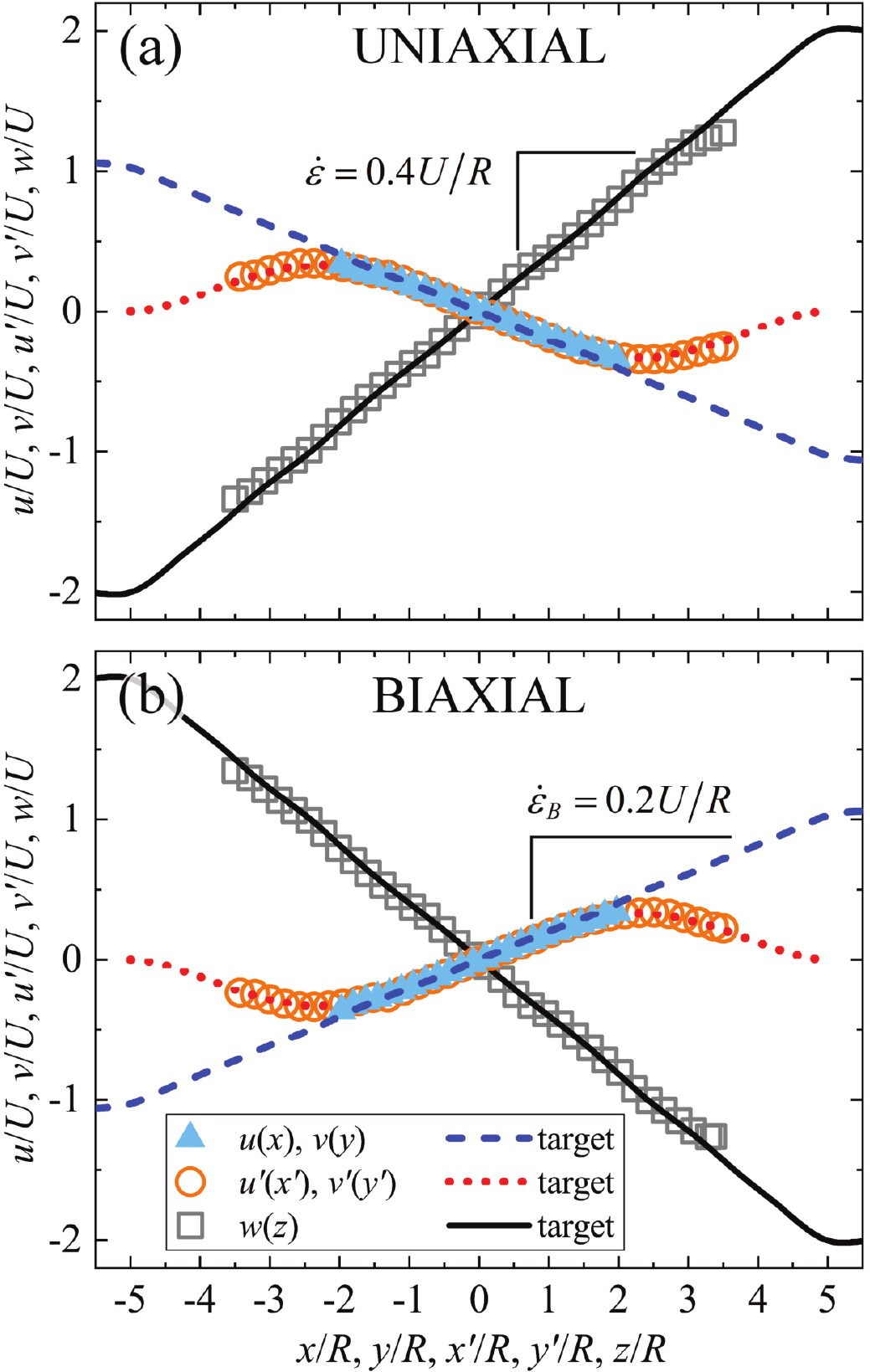}
\caption {Normalized experimental velocity profiles (data points) compared with the target numerical predictions (lines) for (a) uniaxial, and (b) biaxial extension in the OUBER device. Experimental data in (a) and (b) are the mean of all the profiles shown in Figs.~\ref{UniProfiles} and \ref{BiProfiles}, respectively, also assuming that $u(x) \equiv v(y)$ and that $u'(x') \equiv v'(y')$.
} 
\label{norm}
\end{center}
\end{figure}

Normalizing streamwise axial velocity components by the average flow velocity $U$, and normalizing distances by the radius $R$ of the circular cross-section inlet/outlet channels, the experimentally-measured axial velocity profiles for different imposed flow rates collapse, as expected for a Newtonian flow at low $\text{Re}$. Mean normalized profiles computed from five measurements made for imposed volumetric flow rates $0.05~\leq~Q~\leq~0.8$~mL~min$^{-1}$, are shown for uniaxial and biaxial extension in Fig.~\ref{norm}(a) and Fig.~\ref{norm}(b), respectively. Note that these profiles are also computed by taking the mean of $u(x)$ and $v(y)$, and of $u'(x')$ and $v'(y')$ under the (reasonable) assumption that the flow along each of those two pairs of orthogonal directions is similar. The data points shown in Fig.~\ref{norm} represent the normalized streamwise velocity profiles measured experimentally along the $x$ and $y$ axes (orange open circles), the $x'$ and $y'$ axes (light blue closed up-triangles), and along the $z$ axis (gray open squares). The lines shown in Fig.~\ref{norm} represent the target velocity profiles (i.e., the solutions of the Newtonian numerical simulations performed in the target flow geometry) along $x$ and $y$ (dark blue dashed line), along $x'$ and $y'$ (red dotted line), and along $z$ (continuous black line). Over the ranges of measurement, the experimental profiles clearly all agree very well with the target numerical solutions. As mentioned above, if the flow were ideally axisymmetric, the profiles of $u(x)$, $v(y)$, $u'(x')$ and $v'(y')$ would all be identical. In Fig.~\ref{norm} we observe that they agree well for $\abs{x}\lesssim R$ and $\abs{x'} \lesssim R$ (within $\approx 5\%$), but they diverge at greater radial distances from the $z$ axis, i.e., towards the perimeter of the circular region on the $z=0$ plane at $L_1 = 5R$. For $\abs{x} = \abs{x'} = 1.5 R$, $u' \approx 0.85u$, and for $\abs{x} = \abs{x'} = 2 R$, $u' \approx 0.75u$. At $\abs{x'}=5R$, $u' = 0$. Accordingly, in uniaxial extension (Fig.~\ref{norm}(a)), the measured extensional rate along the $z$ axis is close to the numerical prediction $\dot\varepsilon = \partial w /\partial z =0.4U/R$, and is approximately uniform over the range $\abs{z} \leq 5R$. In biaxial extension (Fig.~\ref{norm}(b)), the extensional rate over the $z=0$ plane is evidently also given by the numerical prediction $\dot\varepsilon_B = \partial u /\partial x (=\partial v /\partial y) =0.2U/R$, and is approximately uniform over a circular region defined by $x^2 + y^2 \lesssim R^2$ (but is maintained over greater distances of $\approx \pm 5R$ along the $x$ and $y$ axes with which the planar outlet channels are aligned).

\begin{figure*}[!ht]
\begin{center}
\includegraphics[scale=0.75]{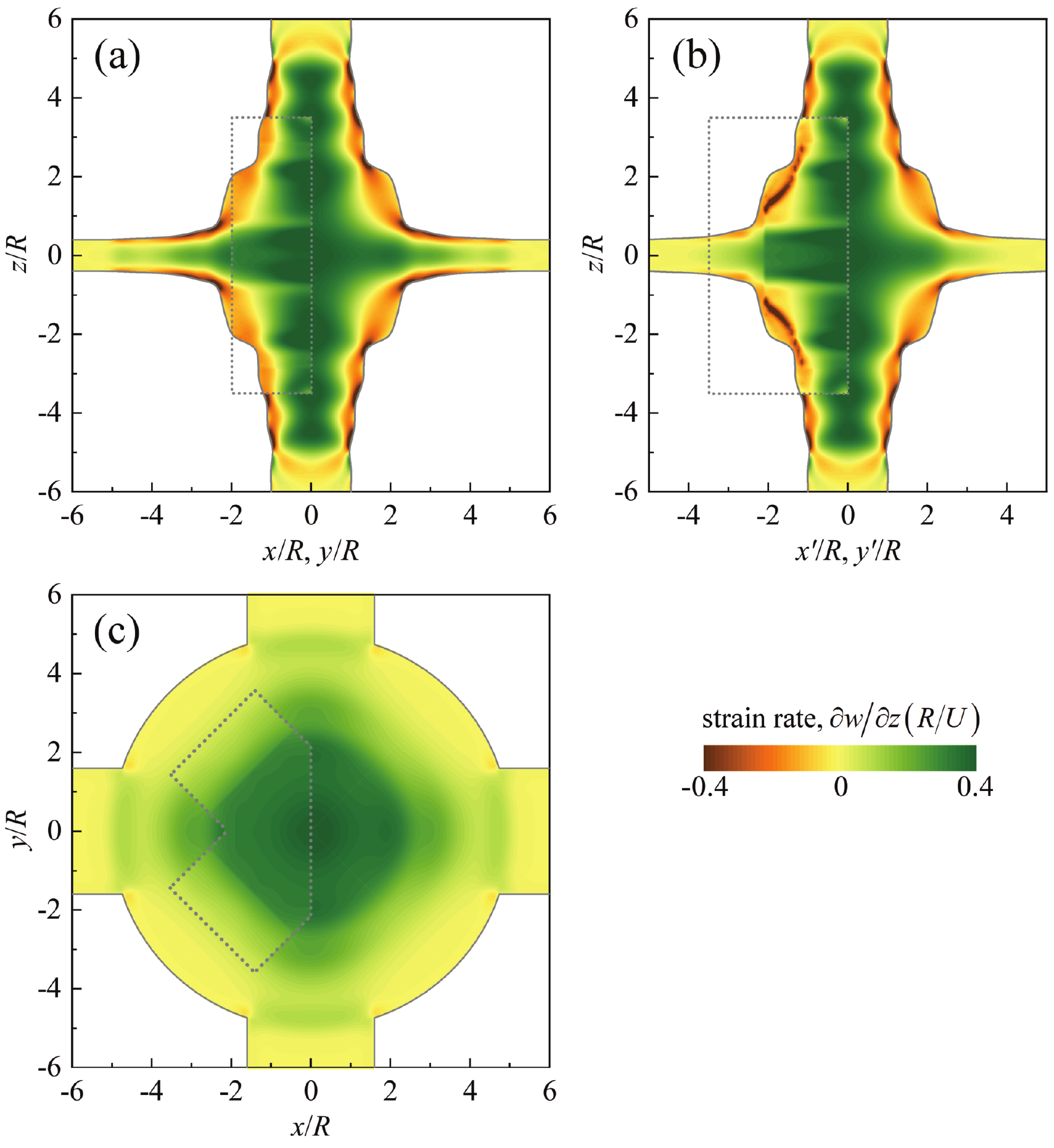}
\vspace{-0.1in}
\caption {Full-field normalized extensional rate for uniaxial extensional flow in the OUBER device $(R/U) \partial w/\partial z$, visualized in (a) the $x=0$ or $y=0$ plane, (b) the $x'=0$ or $y'=0$ plane, and (c) the $z=0$ plane. The result determined from numerical simulation for Newtonian flow is superimposed with the experimental result (shown inside the region marked by gray dotted lines). The experimental result is obtained by averaging mirrored and flipped data obtained for $\text{Re} \approx 0.02$, and by assuming that the planes $x=0$ and $y=0$, and $x'=0$ and $y'=0$, are similar.
} 
\label{Uni_edot}
\end{center}
\end{figure*}

In Fig.~\ref{Uni_edot} we compare the full field numerical prediction and experimental measurement of the extensional rate in uniaxial elongation, $\partial w/\partial z$, normalized by $U/R$. Since the extensional rate is a derived quantity, the experimental result is obtained from an average of mirrored and flipped velocity fields in order to smooth the data. Furthermore, we average data obtained from the $x=0$ and $y=0$ planes, and from the $x'=0$ and $y'=0$ planes, on the assumption that the velocity fields over each of these two pairs of planes should be similar. In Fig.~\ref{Uni_edot}, the experimental result (only available in a limited field of view) is superimposed on the numerical prediction over each imaged plane, and is contained within the boundaries indicated by the dotted gray lines. In all three planes [$x=0$ or $y=0$ (Fig.~\ref{Uni_edot}(a)), $x'=0$ or $y'=0$ (Fig.~\ref{Uni_edot}(b)), and $z=0$ (Fig.~\ref{Uni_edot}(c))] there is an excellent agreement between the numerical prediction and the experimental measurement, with closely matching contours of $\partial w/\partial z$. Also, the normalized extensional rate is close to the expected value of $\dot\varepsilon R/U=0.4$ over the large green regions observed in each plane.

\begin{figure*}[!ht]
\begin{center}
\includegraphics[scale=0.75]{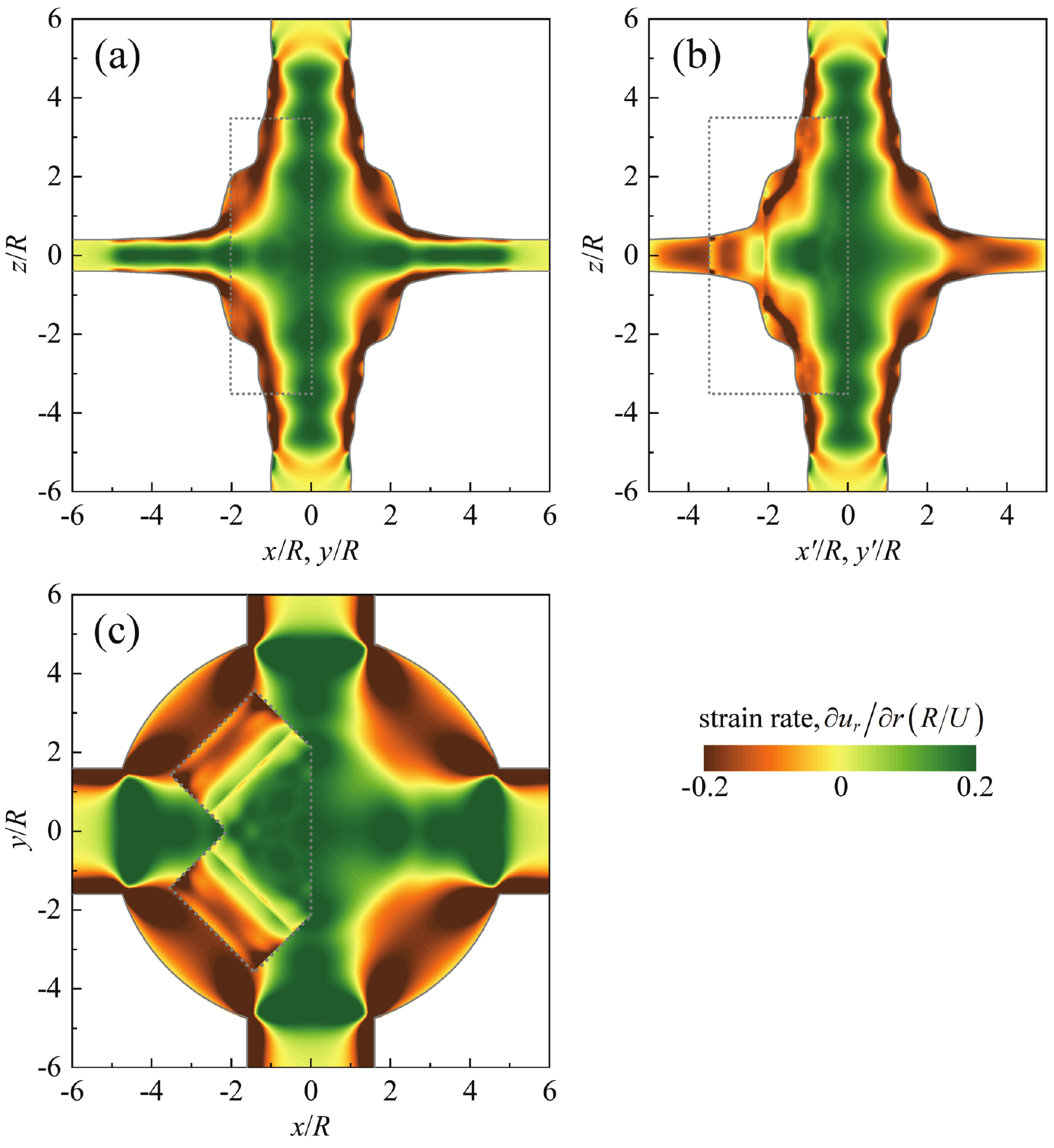}
\vspace{-0.1in}
\caption {Full-field normalized extensional rate for biaxial extensional flow in the OUBER device $(R/U) \partial u_r/\partial r$ (see main text), visualized in (a) the $x=0$ or $y=0$ plane, (b) the $x'=0$ or $y'=0$ plane, and (c) the $z=0$ plane. The result determined from numerical simulation for Newtonian flow is superimposed with the experimental result (shown inside the region marked by gray dotted lines). The experimental result is obtained by averaging mirrored and flipped data obtained for $\text{Re} \approx 0.02$, and by assuming that the planes $x=0$ and $y=0$, and $x'=0$ and $y'=0$, are similar.
} 
\label{Bi_edot}
\vspace{-0.1in}
\end{center}
\end{figure*}

In the same manner as Fig.~\ref{Uni_edot}, a full field comparison between the numerically predicted and experimentally measured extensional rates in biaxial elongation is presented in Fig.~\ref{Bi_edot}. Here, due to the two orthogonal axes of extension (along the $x$ and $y$ directions), we present the data in terms of $\partial u_r/\partial r$, where $u_r = u\cos\theta + v\sin\theta$, $r=\sqrt{x^2 + y^2}$ and $\theta = \tan^{-1}(y/x)$. Once again, the extensional rate is normalized by $U/R$, revealing large green regions where the expected value of $\dot\varepsilon_B R/U=0.2$ is approximated, and showing a generally good agreement between the experiment and the simulation in all three visualized planes [$x=0$ or $y=0$ (Fig.~\ref{Bi_edot}(a)), $x'=0$ or $y'=0$ (Fig.~\ref{Bi_edot}(b)), and $z=0$ (Fig.~\ref{Bi_edot}(c))]. From these full field visualizations of $\partial u_r/\partial r$, the difference between the $x$ and $x'$ (and between the $y$ and $y'$) directions is clearly evident, particularly from the view in the $z=0$ plane (Fig.~\ref{Bi_edot}(c)), where the expected strain rate is maintained for $\approx 5R$ along $x$ and $y$, but becomes negative after $\approx 2R$ along $x'$ and $y'$ as the flow approaches the perimeter of the circular expansion region. The homogeneity of the flow field could very likely be improved by including four additional planar inlet/outlet channels aligned along the positive and negative $x'$ and $y'$ directions. However, this would add complexity to the experimental operation of the device, requiring additional syringes and pumps to control the flow and also further limiting the optical access for any desired quantification of the flow field.

\begin{figure*}[!ht]
\begin{center}
\includegraphics[scale=0.75]{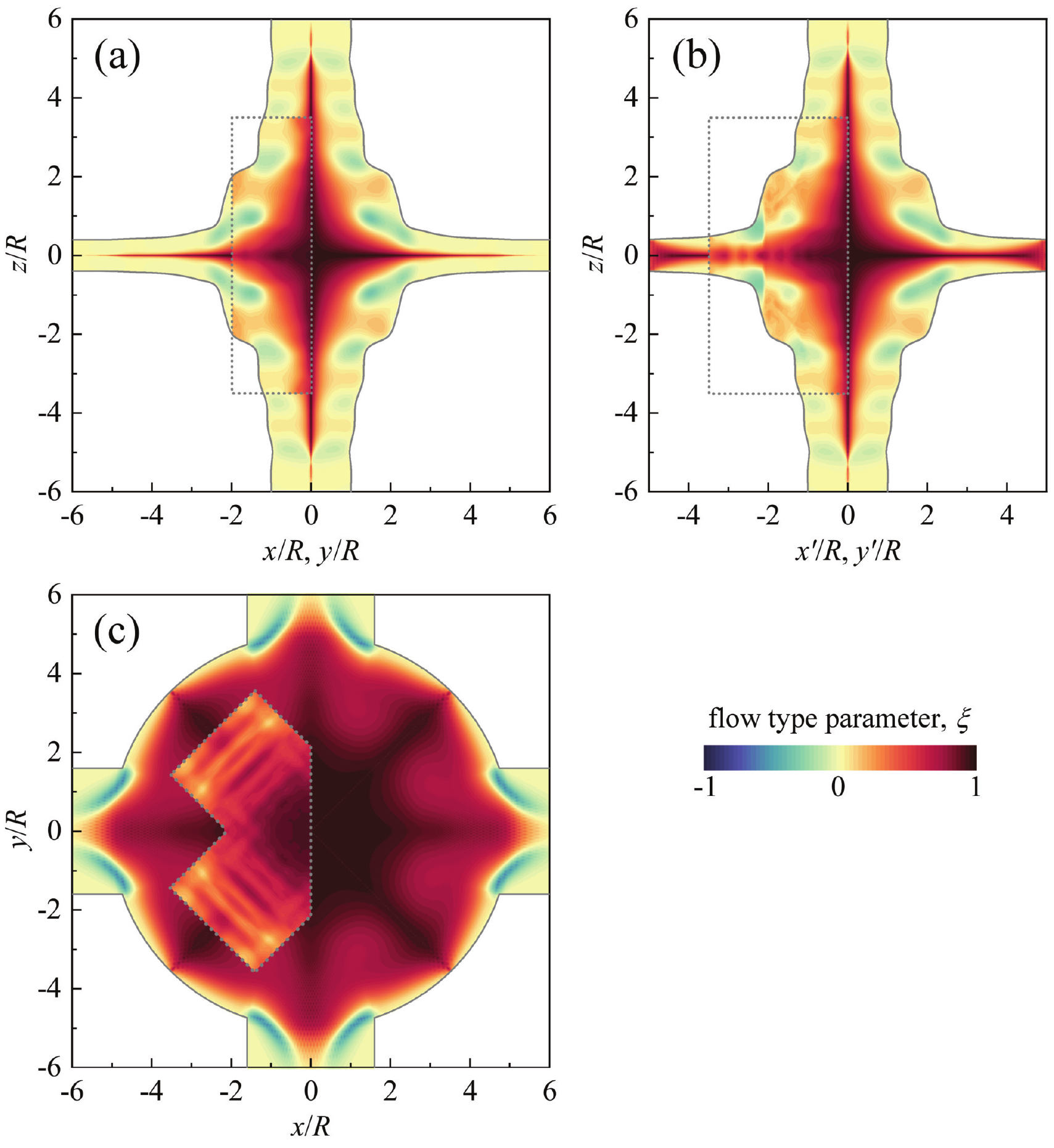}
\vspace{-0.1in}
\caption {Full-field flow type parameter $\xi$ visualized in (a) the $x=0$ or $y=0$ plane, (b) the $x'=0$ or $y'=0$ plane, and (c) the $z=0$ plane. The result determined from numerical simulation for Newtonian flow is superimposed with the experimental result (shown inside the region marked by gray dotted lines). Due to the reversibility of the flow, the flow type parameter is the same in both uniaxial and biaxial extension. The experimental result is obtained by averaging mirrored and flipped data obtained in both flow configurations ($\text{Re} \approx 0.02$) and by assuming that the planes $x=0$ and $y=0$, and $x'=0$ and $y'=0$, are similar.
} 
\label{FTP}
\vspace{-0.1in}
\end{center}
\end{figure*}

Finally, to indicate the local flow kinematics, in Fig.~\ref{FTP} we present a comparison between the full field flow type parameter determined by simulation and experiment. The flow type parameter $\xi = (\abs{\pmb{\dot\upgamma}} - \abs{\pmb{\Omega}} )/( \abs{\pmb{\dot\upgamma}} + \abs{\pmb{\Omega}})$, where $\abs{\pmb{\dot\upgamma}}~=~\sqrt{ \pmb{\dot\upgamma}:\pmb{\dot\upgamma}/2}$ is the magnitude of the deformation rate tensor, $\pmb{\dot\upgamma}$, and $\abs{\pmb{\Omega}}~=~\sqrt{\pmb{\Omega}:\pmb{\Omega}/2}$ is the magnitude of the vorticity tensor, $\pmb{\Omega}$. \cite{Astarita1979} Here, $\xi = -1$ indicates solid body rotation, $\xi = 0$ indicates simple shear, and $\xi = 1$ indicates purely extensional kinematics. Due to the kinematic reversibility of the uniaxial and the biaxial flow configurations, the flow type parameter is expected to be the same for both (indeed the results obtained from the numerical simulations are identical). Therefore, in this case we only show one set of fields and the experimental result shown in Fig.~\ref{FTP} is obtained by averaging the data from uniaxial and biaxial flow. Over the $x=0$ (or $y=0$) plane (Fig.~\ref{FTP}(a)) and over the $x'=0$ (or $y'=0$) plane (Fig.~\ref{FTP}(b)), there is an excellent agreement between the simulation and the experiment, with a very satisfactory matching between countours of $\xi$. In the $z=0$ plane (Fig.~\ref{FTP}(c)), the match between experiment and simulation is less impressive, with the experiment showing a reduced value of $\xi$ in comparison to the simulation. However, it must be remembered that the experimental result is derived from rather heavily smoothed and processed primary data. Also, velocimetry data on the $z=0$ plane has a somewhat low signal to noise ratio, lying directly along the line of sight into the flow cell and having a lower velocity magnitude than most of the field (see Figs.~\ref{UniPIV} and \ref{BiPIV}).

In general, it is evident that the regions of extensionally-dominated flow kinematics (red regions where $\xi \rightarrow 1$ in Fig.~\ref{FTP}) correspond with the regions of approximately uniform extensional rates in uniaxial and biaxial flow (green regions where $\partial w/\partial z \approx 0.4$ in Fig.~\ref{Uni_edot} and where $\partial u_r / \partial r \approx 0.2$ in Fig.~\ref{Bi_edot}, respectively). In short, the OUBER geometry successfully generates regions of nearly pure extensional flow at approximately uniform extensional rate that extend over several characteristic device lengthscales in all three spatial directions. Taken as a whole, the $\upmu$-TPIV of the Newtonian flow field provides a clear confirmation that the fabricated OUBER device closely reproduces the numerically-predicted flow fields, and therefore has potential for use as a uniaxial and biaxial extensional rheometer.

\section{Summary and Conclusions}
\label{SumCon}

We have presented a numerical optimization of the ``6-arm cross-slot'' device~\cite{Afonso2010,Haward2019b} aimed at obtaining a geometry able to impose homogeneous uniaxial and biaxial stagnation point extensional flow fields with the intention of developing a uni- and biaxial extensional rheometer for mobile complex fluids. The optimization procedure (based on solving the Newtonian flow field) yielded a number of different geometries that depended on the input design parameters (i.e., the lengthscales over which the flow field was optimized). Of the generated geometries, one shape in particular was considered most amenable to fabrication and experimental verification of its performance. Prior to the fabrication, it was confirmed by numerical simulations with the Oldroyd-B and l-PTT models that the optimal flow field would also apply to rheologically complex constant viscosity and shear thinning viscoelastic fluids.

The device fabrication itself was achieved at microfluidic dimensions by the technique of selective laser-induced etching of fused silica glass. The fabrication yielded a highly precise match to the numerically-designed geometry in a transparent substrate with optical access to the stagnation point region. Microtomographic particle image velocimetry for flow of a refractive index-matched Newtonian fluid at low Reynolds number was used to quantify the flow field in the experimental geometry over a relatively large volume centered on the stagnation point. These experiments provided confirmation of the good performance of the device.

In conclusion, we have designed, fabricated and thoroughly tested a complex three-dimensional stagnation point microfluidic device, which has shown to produce good approximations to ideal uniaxial and ideal biaxial extension over multiple characteristic lengthscales in each spatial dimension. The applied extensional rate scales linearly with the imposed volumetric flow rate, while the presence of the stagnation point means that the high (infinite) fluid strains requisite for steady-state extensional rheological measurements are achievable. Furthermore, the microfluidic dimensions of the device minimize inertia, which is essential for performing valid extensional viscosity measurements.~\cite{Dontula1997} 

In Part II of this paper,~\cite{Haward2023} we will demonstrate the use of pressure drop measurements in our new OUBER device for extracting the extensional rheological properties of viscoelastic fluids in uniaxial and biaxial extension. Furthermore, in combination with measurements made in the planar OSCER device,~\cite{Haward2012c} we will present a comparison between the uniaxial, planar and biaxial extensional rheometry of model dilute polymeric solutions.

\begin{acknowledgments}
S.J.H, S.V., D.W.C., K.T-P and A.Q.S. gratefully acknowledge the support of the Okinawa Institute of Science and Technology Graduate University (OIST) with subsidy funding from the Cabinet Office, Government of Japan, along with funding from the Japan Society for the Promotion of Science (JSPS, Grant Nos. 21K14080, 21K03884, and 22K14184). F.P. and M.A.A. acknowledge the support provided by LA/P/0045/2020 (ALiCE), UIDB/00532/2020 and UIDP/00532/2020 (CEFT), funded by national funds through FCT/MCTES (PIDDAC). We are indebted to Prof. Robert J.  Poole (University of Liverpool) for insightful discussions.
\vspace{-0.1in}
\end{acknowledgments}

\section*{Data Availability Statement}

The data that support the findings of this study are available from the corresponding author upon reasonable request.

\appendix

\section{Meshes and mesh dependence study}
\label{appendix}

In this appendix, we briefly outline the mesh dependence study carried out to ensure the results of the viscoelastic flow simulations presented in Sec.~\ref{Viscoelastic} are mesh independent. In all simulations we use tetrahedral elements. To check the mesh convergence of our numerical solutions we used three consecutively refined meshes, whose characteristics are quoted in Table~\ref{Table2}. Mesh M3 was used in all other simulations. The coarsest mesh (M1) is represented in Fig.~\ref{mesh}. 

\renewcommand{\thetable}{A\arabic{table}}

\setcounter{table}{0}

\begin{table}[!ht]
\caption{\label{Table2}Characteristics of the numerical meshes used in this study. Element size is that at the stagnation point, normalized by $R$.  }
\begin{ruledtabular}
\begin{tabular}{c c c c }
Mesh   &   \# of elements  &   \# of nodes     &  element size   \\
\hline
M1 & 247830 & 50283 & 0.040           \\
M2 & 713170 & 144193 & 0.028     \\
M3 & 2392852 & 480560 & 0.02   \\

\end{tabular}
\end{ruledtabular}
\end{table}

\renewcommand{\thefigure}{A\arabic{figure}}

\setcounter{figure}{0}
\begin{figure}[!ht]
\begin{center}
\includegraphics[scale=0.32]{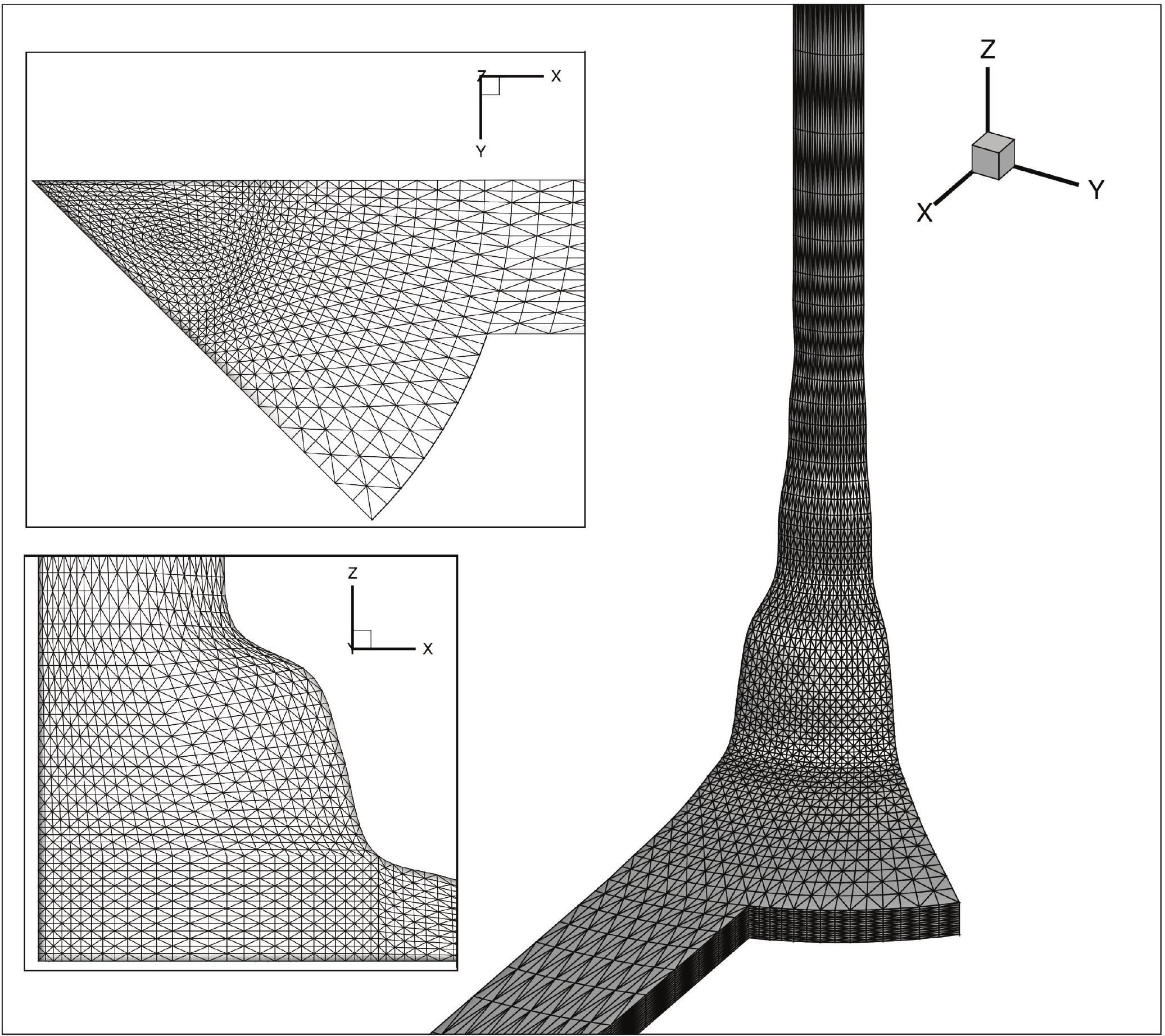}
\vspace{-0.1in}
\caption {Representation of mesh M1 (Table~\ref{Table2}).
} 
\label{mesh}
\vspace{-0.1in}
\end{center}
\end{figure}

\begin{figure}[!ht]
\begin{center}
\includegraphics[scale=0.7]{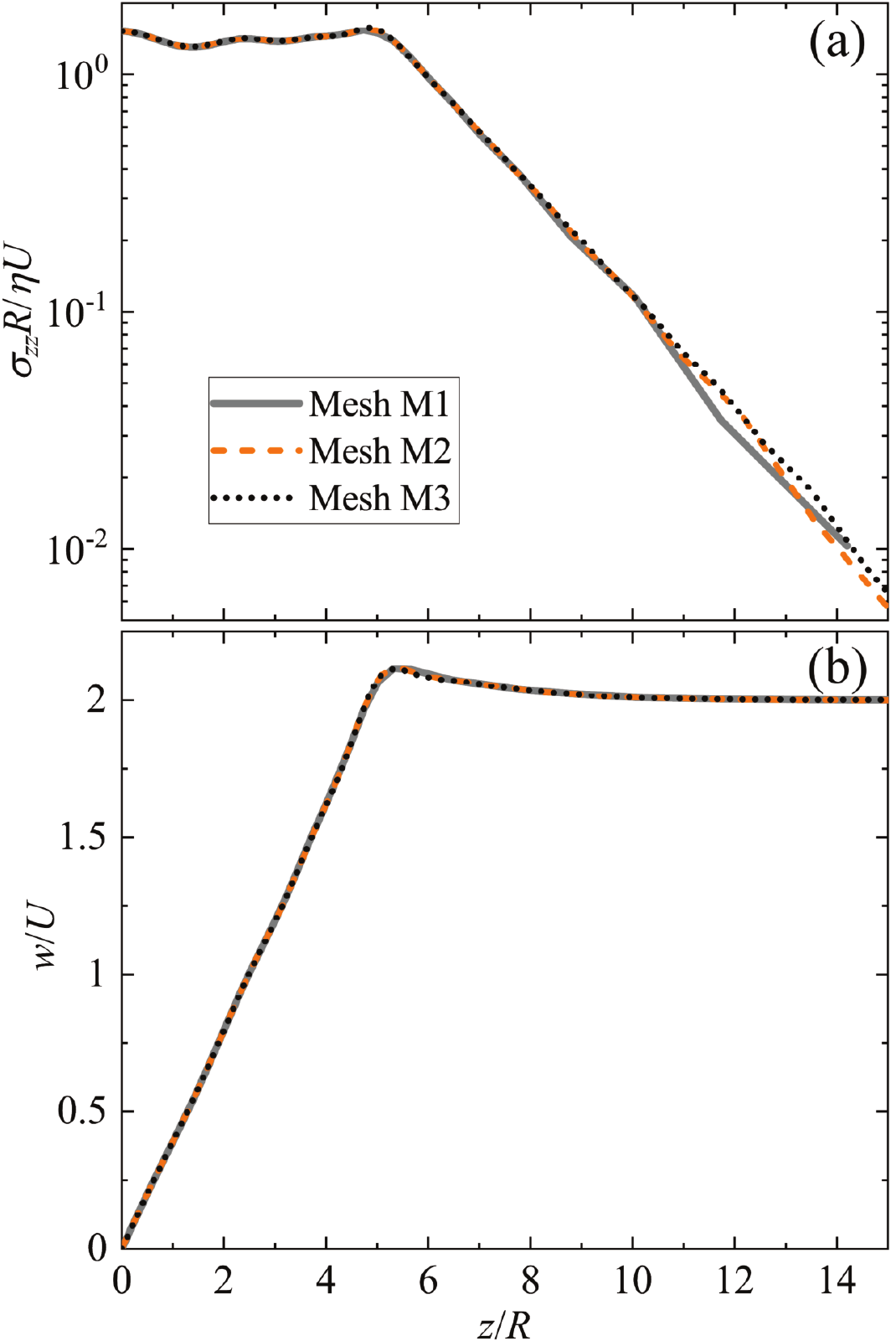}
\vspace{-0.1in}
\caption {Demonstration of mesh independence for the simulation of uniaxial extensional flow in the OUBER geometry with the Oldroyd-B model at $\text{Wi}=0.4$ and with $\beta=0.11$: (a) non-dimensional axial stress $\sigma_{zz}R/\eta U$, and (b) non-dimensional axial velocity $w/U$ as a function of the axial location $z/R$. 
} 
\label{meshindep}
\vspace{-0.1in}
\end{center}
\end{figure}

Fig.~\ref{meshindep} clearly demonstrates the mesh independence for the case of uniaxial flow in the OUBER geometry with the Oldroyd-B model under conditions of $\text{Wi}=0.4$ and with $\beta~=~0.11$. Obtaining mesh independent solutions in this case is the most challenging out of all the cases examined in Sec.~\ref{Viscoelastic}.

\section*{References}

%\begin{thebibliography}
%\bibliography{reference}% Produces the bibliography via BibTeX.
%\end{thebibliography}


\end{document}